\documentclass{IEEEtran}
\usepackage{graphicx}
\usepackage{amsmath}
\usepackage{multicol}
\usepackage{multirow}
\usepackage{stfloats}
\usepackage{booktabs}
\usepackage{mathrsfs}
\usepackage{enumerate}
\usepackage{amssymb}
\usepackage{algorithm}
\usepackage{array}
\usepackage{underscore}
\usepackage{url}
\usepackage{hyperref}
\hypersetup{hidelinks}

\usepackage{flushend}

\usepackage{caption}
\usepackage{subcaption}

\usepackage{color}

\usepackage[outerbars,color]{changebar}
\ifx\pdfoutput\undefined
\else\ifnum\pdfoutput>0
  \usepackage{pdfcolmk}
\fi\fi
\cbcolor{red}
\usepackage{fancyhdr}

\usepackage{algorithm}
\usepackage{algpseudocode}
\usepackage{epstopdf}

\makeatletter
\newcommand{\Rmnum}[1]{\expandafter\@slowromancap\romannumeral #1@}
\makeatother

\usepackage{array}  \newcommand{\PreserveBackslash}[1]{\let\temp=\\#1\let\\=\temp}  \newcolumntype{C}[1]{>{\PreserveBackslash\centering}p{#1}}  \newcolumntype{R}[1]{>{\PreserveBackslash\raggedleft}p{#1}}  \newcolumntype{L}[1]{>{\PreserveBackslash\raggedright}p{#1}}

\ifCLASSINFOpdf
\else
\fi
\begin{document}
\title{\begin{Huge}Adaptive Resonant Beam Charging for \\ Intelligent Wireless Power Transfer\end{Huge}}

\author{Qingqing~Zhang,
Wen~Fang,
Mingliang~Xiong,
Qingwen~Liu\IEEEauthorrefmark{1},
Jun~Wu,
and~Pengfei~Xia

\thanks{
The material in this paper was presented in part at the 2017 IEEE 86th Vehicular Technology Conference (VTC2017-Fall), Toronto, Canada, September 24-27, 2017. 
}

\thanks{Q. Zhang, W. Fang, M. Xiong, Q. Liu, J. Wu, P. Xia, are with Department of Computer Science and Technology, Tongji University, Shanghai, People's Republic of China (e-mail: anne@tongji.edu.cn, wen.fang@tongji.edu.cn, xiongml@tongji.edu.cn, qliu@tongji.edu.cn, wujun@tongji.edu.cn, pengfei.xia@gmail.com).}

\thanks{* Corresponding author.}

\thanks{Copyright (c) 2012 IEEE. Personal use of this material is permitted. However, permission to use this material for any other purposes must be obtained from the IEEE by sending a request to pubs-permissions@ieee.org.}
}

\maketitle

\begin{abstract}
As a long-range high-power wireless power transfer (WPT) technology, resonant beam charging (RBC) can transmit Watt-level power over long distance for the devices in the internet of things (IoT). Due to its open-loop architecture, RBC faces the challenge of providing dynamic current and voltage to optimize battery charging performance. In RBC, battery overcharge may cause energy waste, thermal effects, and even safety issues. On the other hand, battery undercharge may lead to charging time extension and significant battery capacity reduction. In this paper, we present an adaptive resonant beam charging (ARBC) system for battery charging optimization. Based on RBC, ARBC uses a feedback system to control the supplied power dynamically according to the battery preferred charging values. Moreover, in order to transform the received current and voltage to match the battery preferred charging values, ARBC adopts a direct current to direct current (DC-DC) conversion circuit. Relying on the analytical models for RBC power transmission, we obtain the end-to-end power transfer relationship in the approximate linear closed-form of ARBC. Thus, the battery preferred charging power at the receiver can be mapped to the supplied power at the transmitter for feedback control. Numerical evaluation demonstrates that ARBC can save 61\% battery charging energy and 53\%-60\% supplied energy compared with RBC. Furthermore, ARBC has high energy-saving gain over RBC when the WPT is unefficient. ARBC in WPT is similar to link adaption in wireless communications. Both of them play the important roles in their respective areas.

\end{abstract}


\IEEEpeerreviewmaketitle


\section{Introduction}\label{Section1}
Internet of things (IoT) takes significant strides and has become the driving force of scientific revolution and industrial transformation \cite{ding2014cognitive}. However, IoT development faces the challenge of device power endurance. Meanwhile, dramatic growth of the multimedia process in mobile devices leads to significant energy consumption \cite{multimediasignals,yu2014survey}. Carrying a power cord and looking for power supply cause inconvenience for users. Hence, wireless power transfer (WPT) technology becomes an attractive solution for the power hunger \cite{wirelesspathent, wirelesspaper1, wirelesspaper2}.

\begin{figure}
	\centering
	\includegraphics[scale=0.22]{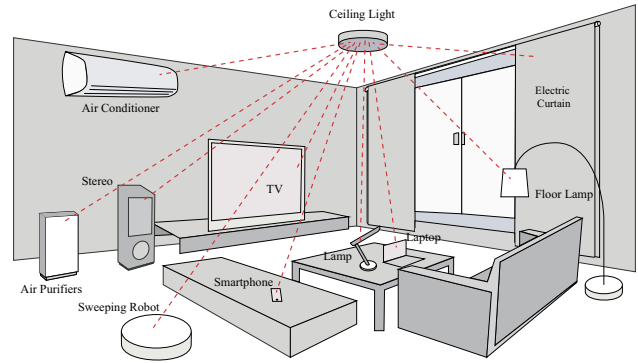}
	\caption{An Example Application Scenario}
	\label{adlcnetwork}
\end{figure}

To provide perpetual power supply for mobile devices, WPT should provide high power over long distance. The existing WPT technologies include inductive coupling, magnetic resonance, radio frequency and laser. However, the transmission distances of inductive coupling and magnetic resonance are only within millimeter or centimeter, which can not support long distance WPT \cite{inductive,magnetic}. Radio frequency and laser are unsafe when transmitting Watt-level power \cite{radio,laser}.

To support Watt-level power transmission over meter-level distance safely, Wi-Charge has published the wireless power delivery products and has received the FDA safety approval \cite{wicharge}. The Wi-Charge transmitter can deliver up to 3 Watts power  over 5 meters to multiple receivers simultaneously through infrared beams while guaranteeing the safety and mobility \cite{wicharge,yu2017internet}. The resonant beam charging (RBC), i.e. distributed laser charging (DLC), was at first presented in \cite{liu2016dlc}.

The RBC mechanism, mathematical models and features are depicted in \cite{dlcvtc,dlciot}. The self-aligning feature of RBC provides users a convenient way to charge their devices without specific aiming or tracking, as long as the transmitter and the receiver are in the line-of-sight (LOS) of each other. RBC supports charging multi-device simultaneously, which is like multi WiFi-devices connecting to a single access point\cite{wicharge,dlcvtc,dlciot}. Additionally, once there is an obstacle between the RBC transmitter and receiver, WPT can be cut off right away. Therefore, the RBC system guarantees safety.

On the other hand, to keep all IoT devices accessing to the RBC system working as long as possible for fairness, the first access first charge (FAFC) scheduling algorithm was presented in \cite{fafc}. Multi-device can be selected to charge with their batteries' preferred charging power according to the accessing order, while all receivers discharge depending on their using statues during a time slot.

As smart-home has become an important IoT application area, Fig.~\ref{adlcnetwork} gives an example of RBC application in an indoor scenario. In Fig.~\ref{adlcnetwork}, the ceiling light is the RBC-equipped light bulb where a RBC transmitter is embedded in. All the electronic devices embedded with the RBC receivers in the coverage of the RBC transmitter can be charged wirelessly and simultaneously.

\begin{figure}
	\centering
	\includegraphics[scale=0.4]{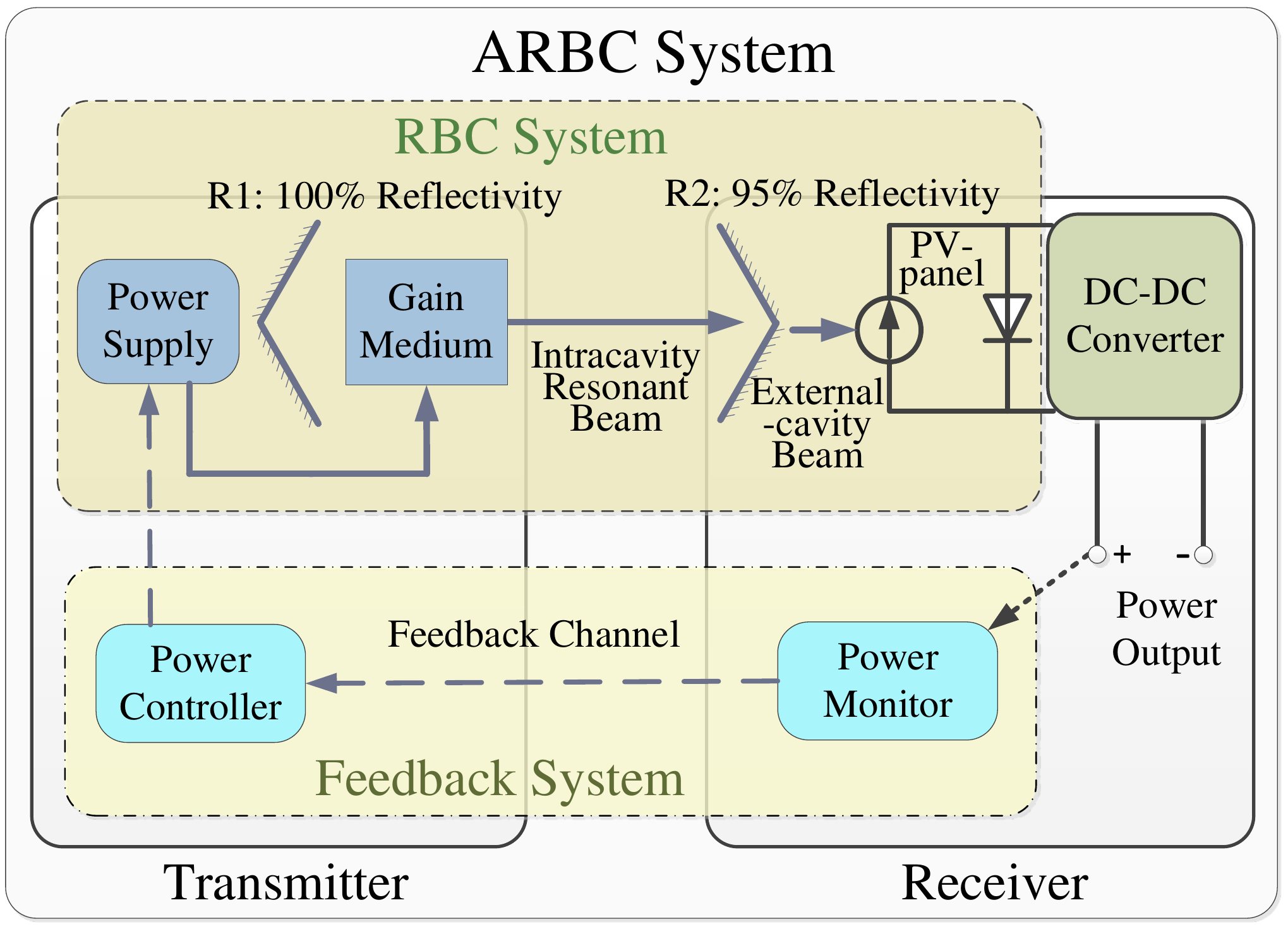}
	\caption{Adaptive Resonant Beam Charging System Diagram}
	\label{adaptivepower}
\end{figure}

However, the preferred charging current, voltage, and thus power of  battery, such as the most widely used Li-ion battery, keep changing during the charging period \cite{goodenough2013li,shen2012charging}. If the transmitter could adjust the emitted power according to the battery preferred value, battery charging performance and wireless power transfer efficiency can be optimized. To this end, an adaptive resonant beam charging (ARBC) system based on the RBC system is introduced in this article.

The contributions of this paper can be summarized as follows: 1) we propose the ARBC system design, which can automatically adjust the resonant beam power relying on the feedback control;
2) we illustrate the working flow, create the mathematical model, and present the control algorithm of the ARBC system, which provide the guidelines for the ARBC system implementation;
3) we analyze the ARBC system performance considering various conditions including beam wavelength, PV-cell temperature, air quality, etc.. We find 61\% battery charging energy and 53\%-60\% supplied energy can be saved  by the ARBC system compared with the RBC system.

In this paper, RBC will be briefly introduced at first. Then, we will introduce the ARBC system. The modules and working mechanisms of the ARBC system will be depicted in the following. The ARBC system design, including the mathematical modeling and the control algorithm, will be presented to quantitatively analyze the ARBC model. Based on the system design, we will evaluate the system performance by using MATLAB and Simulink. Finally, the conclusions will be given and the open issues for future research will be discussed.

\section{System Architecture}\label{Section2}
In this section, we will briefly introduce the RBC system at first \cite{liu2016dlc}. Then, we will propose the ARBC system based on the RBC system.

\subsection{RBC System}\label{}
Fig.~\ref{adaptivepower} shows the RBC system, where the transmitter and the receiver are separated in space. A power supply, a retro-reflector R1 with 100\% reflectivity and a gain medium, which is used to amplify photons, are included in the RBC transmitter. While in the RBC receiver, a retro-reflector R2 with 95\% reflectivity is contained. A photoelectric conversion component, such as photovoltaic-panel (PV-panel), is installed behind R2 at the receiver.

In the RBC system, the electrical power provided by the power supply is converted to the intra-cavity resonant beam power. Then, the intra-cavity resonant beam travels through the air from the transmitter to the receiver with certain attenuation. At the receiver, the intra-cavity resonant beam can be partially converted to the external-cavity beam after passing through R2. Then, the PV-panel converts the external-cavity beam power to the electrical power. Thus, batteries can be charged by the electrical power.

\subsubsection{Electricity-to-Beam Conversion}\label{}
The power supply provides electrical power $P_s$ to pump the gain medium. $P_s$ depends on the stimulating current $I_t$ and voltage $V_t$ as:
\begin{equation}\label{ps}
P_s = I_{t} V_{t}.
\end{equation}

Then, the intra-cavity resonant beam can be stimulated out from the gain medium. If there is no transmission attenuation, the external-cavity resonant beam power at the receiver $P_{bt}$ can be derived as \cite{laserdiodes}:
\begin{equation}\label{powervscurrent}
P_{bt}=\gamma \frac{h\nu}{q}(I_{t}-I_{th}),
\end{equation}
where $\gamma$ is the modified coefficient, $h$ is the Plunk constant, $\nu$ is the beam frequency, $q$ is the electronic charge constant, and $I_{th}$ is the current threshold.

\subsubsection{Beam Transmission}\label{}
The intra-cavity resonant beam travels through the air and arrives at the RBC receiver. During the transmission, the beam power suffers from attenuation, which is similar to electromagnetic (EM) wave propagation power loss \cite{attenuation}.

The beam transmission efficiency $\eta_{bt}$ can be modeled as \cite{JMLiuphotonic}:
\begin{equation}\label{etalt}
  \eta_{bt}=\frac{P_{br}}{P_{bt}}= e^{-\sigma R},
\end{equation}
where $P_{br}$ is the received external-cavity beam power at the receiver, $\sigma$ is the beam attenuation coefficient, and $R$ is the transmission radius. When $R$ is close to zero, $\eta_{bt}$ approaches 100\%, and thus $P_{br}$ is approximate to $P_{bt}$.

\subsubsection{Beam-to-Electricity Conversion}\label{}
At the receiver, the external-cavity beam power $P_{br}$ can be received by a PV-panel and then be converted to electrical power \cite{edouard2013mathematical,aziz2014simulation}. The relationship between the PV-panel output current $I_{o,pv}$ and voltage $V_{o,pv}$ can be depicted as:
\begin{equation}\label{Iopv}
  I_{o,pv}=I_{sc} - I_{s}(e^{V_{o,pv}/V_m}-1),
\end{equation}
where $I_{sc}$ is the PV-panel short-circuit current, and $I_s$ is the saturation current, i.e., the diode leakage current density in the absence of light. $V_m$ is the ``thermal voltage'', which can be defined as:
\begin{equation}\label{Vm}
  V_m=\frac{nkT}{q},
\end{equation}
where $n$ is the PV-panel ideality factor, $k$ is the Boltzmann constant, and $T$ is the PV-cell temperature. Then, the PV-panel output power $P_{o,pv}$ can be obtained as:
\begin{equation}\label{popv}
  P_{o,pv}=I_{o,pv}V_{o,pv}.
\end{equation}

In the RBC system, the electrical power $P_{o,pv}$ can finally be used to charge batteries.


\subsection{ARBC System}\label{}
The RBC system can transmit Watt-level power over long distance \cite{liu2016dlc}. However, there exist some concerns with the RBC system:

1) To optimize battery charging performance, the battery charging current and voltage should be dynamically changed during the charging period, which will be discussed in detail in the following. However, the RBC system can only charge batteries with constant current and voltage.

2) The RBC resonant beam power propagation loss depends on transmission radius and medium obstacle along the LOS path between the transmitter and the receiver \cite{liu2016dlc}. Thus, the space-time varying propagation loss requires dynamic power supply compensation.

3) If the PV-panel output power at the RBC receiver is not fully converted to the battery power, the extra energy usually causes thermal effects, which may lead to PV-panel conversion efficiency reduction, battery damage, and even danger.

To deal with these issues, an intuitive idea is to control the transmission power by adaptively sending the value of battery preferred power from the receiver to the transmitter through a feedback channel. The similar mechanism for signal transmission is well-known as link adaption in wireless communication for optimizing the information delivery \cite{ding2009ofdm}.

At first, we need to specify the adaption goal to optimize battery charging performance. Battery charging profile is the algorithm of using dynamic current and voltage in battery charging process. Battery performance, in terms of achievable capacity and charging speed, depends on the battery charging profile. For example, the preferred battery charging profile of Li-ion battery is known as the constant current - constant voltage (CC-CV) profile to achieve the maximum battery capacity \cite{shen2012charging}.

On the other hand, the PV-panel output current and voltage may not be the battery preferred charging values, the DC-DC converter can take the role of converting the PV-panel output values to the battery preferred ones.

By incorporating the feedback system and the DC-DC converter into the RBC system, the ARBC system architecture can be depicted as in Fig.~\ref{adaptivepower}. The feedback system crosses over the two ends of the ARBC system, which consists of power monitor and power controller.

\begin{figure}
	\centering
	\includegraphics[scale=0.6]{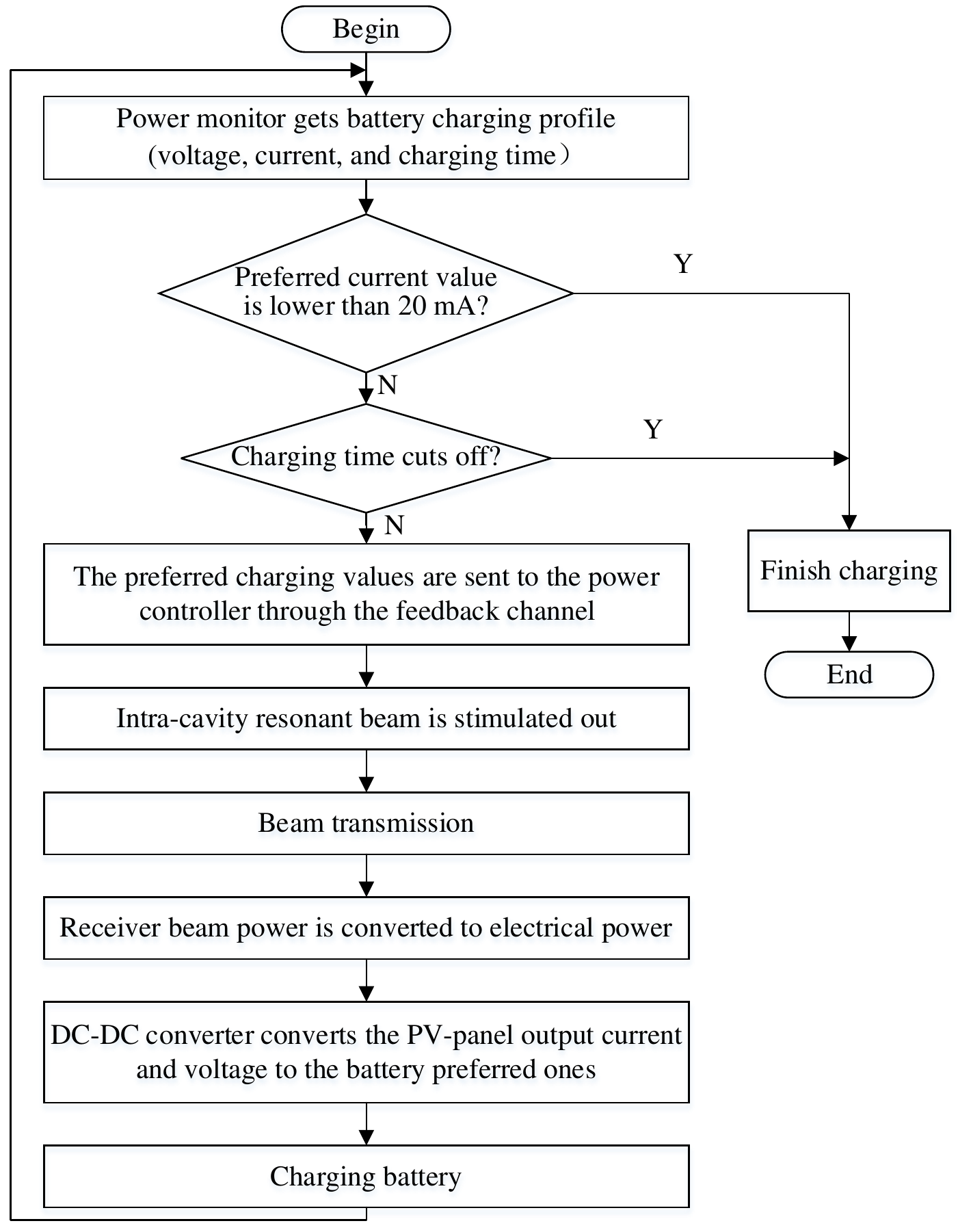}
	\caption{ARBC Flow Chart}
	\label{adaptiveflowchart}
\end{figure}

The ARBC operation flow is depicted in Fig.~\ref{adaptiveflowchart} as:

1) The power monitor gets the values of battery preferred charging current, voltage, and cut-off time;

2) If the preferred charging current is lower than 20 mA, the charging procedure ends. Or, turn to 3);

3) If the charging time cuts off, the charging procedure ends. Or, turn to 4);

4) The power monitor sends the preferred charging power to the power controller;

5) After receiving the preferred charging values, the power controller informs the power supply to generate the corresponding electrical power. The electrical power has effects on the gain medium to stimulate out the intra-cavity resonant beam;

6) The resonant beam travels through the air from the transmitter and arrives at the receiver;

7) The receiver beam power is converted to the electrical power by the PV-panel at the receiver;

8) The DC-DC converter converts the PV-panel output current and voltage to the battery preferred values;

9) The battery is charged with the preferred current and voltage;

10) The power monitor updates the values of battery preferred charging current, voltage, and cut-off time according to the battery status.

Repeating this flow, the battery can be charged with the preferred values during the whole charging procedure.

To specify the ARBC system in detail, the battery charging profile, the DC-DC converter and the feedback mechanism is discussed in the following.

\subsubsection{Li-ion Battery Charging Profile}\label{}
Different kinds of batteries may have different charging profiles given the chemical characteristics \cite{cleveland2008battery}. Li-ion battery is the most widely used rechargeable battery for IoT and mobile devices. In traditional charging systems, including the RBC system, batteries are charged with fixed power. However, for Li-ion battery, even slightly undercharging can lead to significant capacity reduction \cite{dearborn2005charging}. For example, 1.2\% undercharge of the optimal full-charge voltage may result in 9\% capacity reduction. On the other hand, overcharge may damage the battery and even cause danger. Therefore, offering controllable current and voltage is important to charge Li-ion battery safely as well as achieve its full capacity.

We discuss here the Li-ion battery preferred charging profile, which includes four stages as in Fig.~\ref{li-ionchargezero}. As an instance of a single cell Li-ion battery with 1000 mAh capacity, the battery can theoretically provide one hour of operating time when discharged at a constant current of 1000 mA. We outline the stages of battery charging profile as \cite{chargingprofile,dearborn2005charging}:

\begin{figure}
	\centering
	\includegraphics[scale=0.6]{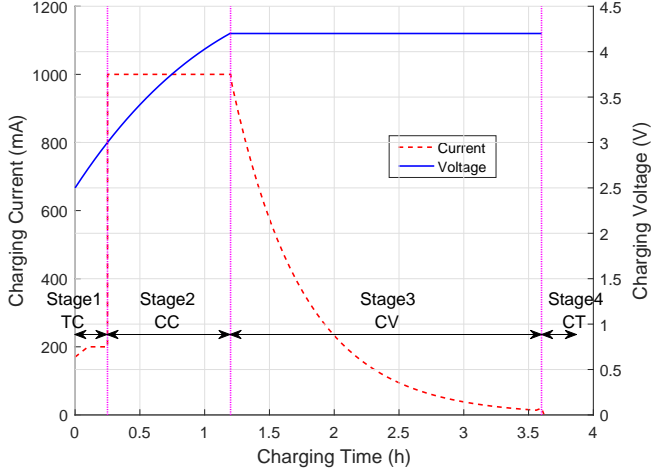}
	\caption{Li-ion Battery Charging Profile}
	\label{li-ionchargezero}
\end{figure}

Stage 1: Trickle Charge (TC) - When the battery voltage is below 3.0 V, the battery is charged with the current which increases towards 200 mA. The voltage rises to 3.0 V.

Stage 2: Constant Current (CC) Charge - After the battery voltage has risen above 3.0 V, the TC-CC threshold, the charging current switches to constant value between 200 mA to 1000 mA. The voltage rises towards 4.2 V.

Stage 3: Constant Voltage (CV) Charge - When the cell voltage reaches the CC-CV threshold, 4.2 V, the CC stage ends and the CV stage begins. In order to maximize capacity, the voltage variation tolerance should be less than ±1\%. The current decreases towards 20 mA.

Stage 4: Charge Termination (CT) –- Two approaches are typically used to terminate charging: 1) minimum current charge or 2) timer cutoff. However, a combination of the two techniques may also be applied. In the minimum current approach, battery charge is terminated when the current diminishes below 20 mA, the minimum current threshold, during the CV stage. In the timer cutoff approach, for example, 2.4-hour timer starts when the CV stage is invoked. The charge is terminated after 3.6-hour during the CV stage.

It takes 4 hours to fully charge a deeply depleted battery to maximize the battery capacity. The charging speed is affected by CC in Stage~2 \cite{shen2012charging}. For example, if CC is 700 mA, 50\% to 70\% capacity can be charged at the end of Stage~2. This is the secret that many ``fast-charging'' techniques rely on. If CC is 200 mA, much longer time is needed to finish Stage~2, however, nearly 100\% capacity can be achieved at the end of Stage~2. It is the tradeoff between charging speed and achieved capacity, which can be controlled by CC at Stage~2.

Compared with fixed-charging system, dynamically charging, according to the above charging profile, can not only avoid overcharge or undercharge, but also reduce potential damage to battery or safety concern.

\subsubsection{DC-DC Conversion}\label{}
The PV-panel output current $I_{o,pv}$ and voltage $V_{o,pv}$ may not be optimal for battery charging. At first, $I_{o,pv}$ and $V_{o,pv}$ may be dynamic due to the variance of $P_{br}$ and PV-panel characteristics. Secondly, the preferred battery charging current and voltage vary with different battery conditions, as discussed above. Therefore, converting $I_{o,pv}$ and $V_{o,pv}$ to the battery preferred values becomes imperative.

\begin{figure}
	\centering
	\includegraphics[scale=0.9]{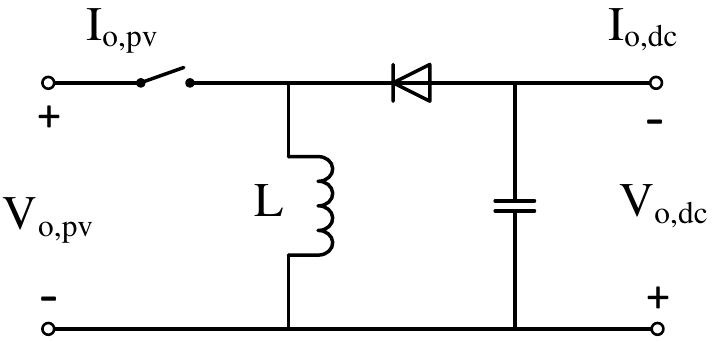}
	\caption{Buck-Boost DC-DC Converter}
	\label{DC-DC}
\end{figure}

In solar power systems, it is well-known to adopt a DC-DC converter between PV-panel and power load to obtain the preferred current and voltage \cite{dancy2000high}. There are three DC-DC converter types -- boost converter, buck converter and buck-boost converter. At the ARBC receiver, the buck-boost converter is adopted \cite{walker2004cascaded}. As depicted in Fig. \ref{DC-DC}, the DC-DC converter, a programmable integrated circuit, can convert the input current and voltage, which are the PV-panel output current $I_{o,pv}$ and voltage $V_{o,pv}$, to the output current $I_{o,dc}$ and voltage $V_{o,dc}$, which are the battery preferred charging current and voltage.

There are two working modes of the buck-boost DC-DC converter according to whether the current through the inductor falls to zero during a working period. If the current through the inductor never falls to zero, it is called the continuous mode. Otherwise, it is the discontinuous mode.

When working at the continuous mode, the relationship between $V_{o,pv}$ and $V_{o,dc}$ can be depicted as:
\begin{equation}\label{dccontinuous}
  \frac{V_{o,dc}}{V_{o,pv}} = - \frac{D}{1-D},
\end{equation}
where $D$ is the duty cycle, which means the switch closing time of the whole working time $t$. While, if working at the discontinuous mode, $V_{o,pv}$ and $V_{o,dc}$ are related as:
\begin{equation}\label{dcdiscontinuous}
  \frac{V_{o,dc}}{V_{o,pv}} = - \frac{{V_{o,pv}} D ^ 2 \ t}{2\ L \ I_{o,pv}},
\end{equation}
where $L$ is the inductor.

Then, the DC-DC converter output current and voltage are converted to the battery preferred charging current and voltage. Thus, batteries can be charged with the preferred values.

\begin{figure}
	\centering
    \includegraphics[scale=0.6]{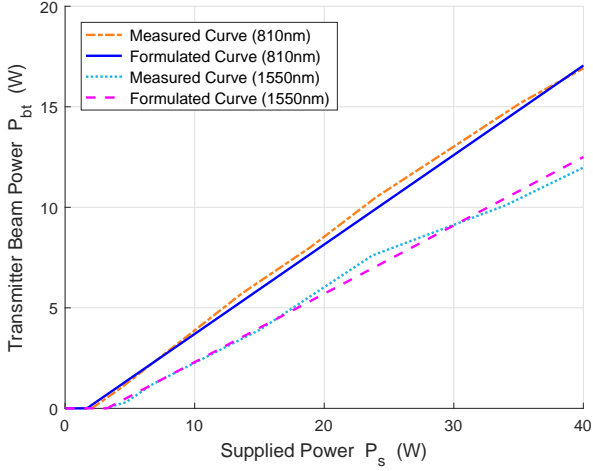}
	\caption{Transmitter Beam Power vs. Supplied Power}
    \label{8101550plps}
\end{figure}

\subsubsection{Feedback Mechanism}\label{}
To charge batteries dynamically and continuously, we should track the preferred battery charging current, voltage, and thus power, and send the information to the power supply during the charging procedure. In the feedback system, the power monitor at the ARBC receiver takes the role of tracking the battery status and obtaining the charging values of current and voltage. After that, the expected charging power can be calculated as the product of the preferred current and voltage.

Then, the adjustment or requirement of the transmitting power will be sent to the power controller by the power monitor through the feedback channel. The feedback channel can be established relying on various wireless communication technologies, e.g., WiFi, Bluetooth, infra-communication. Alternatively, the decision logic can be implemented in the power controller, if the power monitor can feedback the battery charging related information to the power controller.

\begin{table}
\centering
\caption{Curve-fitting Coefficients for Beam Power}
\begin{tabular}{C{2.5cm} C{1.5cm} C{1.5cm}}
\hline
Beam Wavelength &  $a_{1}$ & $b_{1}$  \\
\hline
\bfseries{810 nm} & 0.445 & -0.75 \\
\bfseries{1550 nm} & 0.34 & -1.1 \\
\hline
\label{a1b1}
\end{tabular}
\end{table}

\begin{table}
\newcommand{\tabincell}[2]{\begin{tabular}{@{}#1@{}}#2\end{tabular}}
\centering
\caption{Beam Transmission Parameters}
\begin{tabular}{C{0.6cm} C{6.0cm}}
\hline
 \textbf{Parameter} &\tabincell{c}{\textbf{Value} \\ \quad \textbf{Clear Air} \qquad\quad \textbf{Haze} \qquad\qquad\quad \textbf{Fog}} \\
\hline
\bfseries{$\tau$} & {\qquad \  $10\ km$ \qquad\qquad $3\ km$ \qquad\qquad\  $0.4\ km$} \\
\bfseries{$\theta$}   & {\qquad $1.3$ \qquad\quad $0.16\tau+0.34$ \qquad\quad $0$} \\
\hline
\label{pathloss}
\end{tabular}
\end{table}

Thus, the battery can be charged with preferred current and voltage continuously. Thereafter, the intelligent wireless charging technology can be realized to optimize battery performance. This architecture is capable to support charging different kinds of batteries with diverse charging profiles, such as Li-ion, Ni-MH, etc..

In summary, the proposed ARBC mechanism for optimizing wireless charging performance is similar to the link adaption widely-used in wireless communications for optimizing information delivery.

In the next section, we will present the numerical models and performance evaluation of the ARBC system.


\section{ARBC System Design}\label{Section3}
To design the ARBC system, the power relationship between the battery preferred charging power and the supplied power should be obtained. So, we will introduce the numerical models at first in this section. Based on these models, we will design the system control algorithm to describe the ARBC system in detail.

\subsection{Mathematical Modeling}\label{}
We at first introduce the electricity-to-beam conversion model, the beam transmission model, and the beam-to-electricity conversion model. Based on these models, the ARBC end-to-end power transfer relationship between the battery power and the supplied power can be obtained, which offers a quantitative and intuitive tool to evaluate the ARBC system.

\begin{figure}
	\centering
    \includegraphics[scale=0.6]{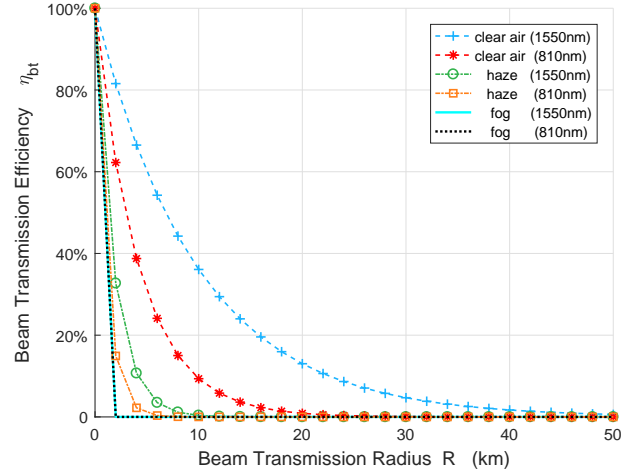}
	\caption{Beam Transmission Efficiency vs. Radius}
    \label{transmittance}
\end{figure}

\subsubsection{Electricity-to-Beam Conversion}\label{}
From \cite{810nmtransmitter,1550nmtransmitter}, the measured values of stimulation current $I_t$, stimulation voltage $V_t$ and the resonant beam power $P_{bt}$  for the beam wavelength of 810 nm and 1550 nm can be obtained, respectively. Then, the supplied electrical power $P_{s}$ can be calculated according to \eqref{ps}. According to \cite{dlcvtc,dlciot}, $P_{s}$ is verified to be linearly related with $P_{bt}$. The relationship between $P_{bt}$ and $P_{s}$ can be described as:
\begin{equation}\label{plps}
P_{bt} \approx a_{1} P_{s}+b_{1},
\end{equation}
where $a_1$ and $b_1$ are two coefficients, of which the values are listed in Table~\ref{a1b1}.

\begin{table}[bp]
\newcommand{\tabincell}[2]{\begin{tabular}{@{}#1@{}}#2\end{tabular}}
\centering
\caption{PV-panel Simulation Parameters}
\begin{tabular}{C{3.25cm} C{4.5cm}}
\hline
 \textbf{Parameter}  & \tabincell{c}{\textbf{Value} \\ \textbf{810 nm} \qquad\qquad \textbf{1550 nm}} \\
\hline
\bfseries{Short-circuit current}                               & \tabincell{c}{$0.16732 A$ \qquad\qquad $0.305 A$\qquad\ }\\
\bfseries{Open-circuit voltage}                              & \tabincell{c}{$1.2 V$ \qquad\qquad $0.464 V$} \\
\bfseries{Irradiance used for measurement}            & \tabincell{c}{$36.5 W/cm^2$ \qquad\qquad $2.7187 W/cm^2$} \\
\bfseries{Laser frequency}                                     & \tabincell{c}{$3.7037\times10^{14} Hz$ \quad $1.9355\times10^{14} Hz$} \\
\bfseries{Quality factor}                                        & \tabincell{c}{$1.5$ \qquad\qquad\quad $1.1\qquad$} \\
\bfseries{Number of series cells}                           & $72$ \\
\bfseries{PV-panel material }                                & \tabincell{c}{GaAs-based \qquad\qquad GaSb-based}  \\
\bfseries{Measurement temperature}                      & \tabincell{c}{$25^{\circ}C$ \qquad\qquad\  $120^{\circ}C$} \\
\hline
\label{simulink}
\end{tabular}
\end{table}

In Fig.~\ref{8101550plps}, the dot-dash-line and the dot-line depict the measured relationships between $P_{bt}$ and $P_{s}$ when the beam wavelength takes 810 nm and 1550 nm, respectively. While, the solid-line and the dash-line are the formulated fitting curves for 810 nm and 1550 nm based on \eqref{plps}. As can be seen, the formulated curves match the measured ones very well, which validates the linear approximation.

\subsubsection{Beam Transmission}\label{}
In different transmission scenarios, the intra-cavity resonant beam power takes different attenuation according to \eqref{etalt}. The attenuation coefficient $\sigma$ can be depicted as \cite{JMLiuphotonic}:
\begin{equation}\label{sigma}
  \sigma = \frac{3.91}{\tau} \Big(\frac{\lambda}{550\ nm}\Big)^{-\theta},
\end{equation}
where $\tau$ is the visibility, and $\theta$ is the size distribution of the scattering particles.

\begin{figure}
	\centering
	\includegraphics[scale=0.6]{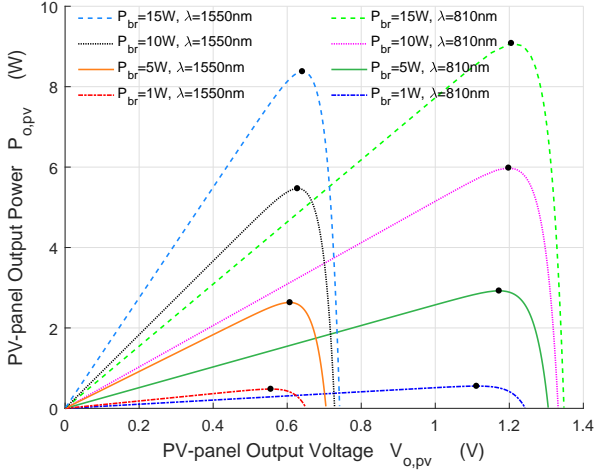}
	\caption{PV-panel Output Power vs. Voltage (25$^\circ$C)}
	\label{8101550Pirradiance}
\end{figure}

\begin{table}[b]
\centering
\caption{Curve-fitting Coefficients for Battery Power}
\begin{tabular}{C{2.1cm} C{1.5cm}  C{1.2cm} C{1.2cm}}
\hline
Beam Wavelength &  Temperature &  $a_{2}$ & $b_{2}$  \\
\hline
\bfseries{810 nm} & {0$^\circ$C} & 0.6084 & -0.08382 \\
\bfseries{ } & {5$^\circ$C} & 0.6087 & -0.08506 \\
\bfseries{ } & {10$^\circ$C} & 0.6089 & -0.08628 \\
\bfseries{ } & {15$^\circ$C} & 0.6092 & -0.08749 \\
\bfseries{ } & {20$^\circ$C} & 0.6094 & -0.08868 \\
\bfseries{ } & {25$^\circ$C} & 0.6096 & -0.08987 \\
\bfseries{ } & {30$^\circ$C} & 0.6098 & -0.09102 \\
\bfseries{ } & {35$^\circ$C} & 0.6100 & -0.09217 \\
\bfseries{ } & {40$^\circ$C} & 0.6102 & -0.09331 \\
\bfseries{ } & {45$^\circ$C} & 0.6103 & -0.09443 \\
\bfseries{ } & {50$^\circ$C} & 0.6105 & -0.09557 \\
\bfseries{1550 nm} & {0$^\circ$C} & 0.6043 & -0.1275 \\
\bfseries{ } & {5$^\circ$C} & 0.5964 & -0.1294 \\
\bfseries{ } & {10$^\circ$C} & 0.5885 & -0.1317 \\
\bfseries{ } & {15$^\circ$C} & 0.5806 & -0.1338 \\
\bfseries{ } & {20$^\circ$C} & 0.5727 & -0.1358\\
\bfseries{ } & {25$^\circ$C} & 0.5649 & -0.1382 \\
\bfseries{ } & {30$^\circ$C} & 0.5569& -0.1398 \\
\bfseries{ } & {35$^\circ$C} & 0.5491 & -0.1424 \\
\bfseries{ } & {40$^\circ$C} & 0.5412 & -0.1440 \\
\bfseries{ } & {45$^\circ$C} & 0.5334 & -0.1464 \\
\bfseries{ } & {50$^\circ$C} & 0.5255 & -0.1483 \\
\hline
\label{a2b2}
\end{tabular}
\end{table}

We consider three typical scenarios, i.e., clear air, haze and fog here. $\theta$ can be specified as \cite{foghaze}:
\begin{equation}\label{rhoc}
\theta = \left\{
             \begin{array}{lr}
             1.3 \qquad\qquad\ \  \mathrm{for \  clear \  air} \ \  (6\ km\leq \tau \leq 50\ km), &  \\
             0.16 \tau+0.34 \ \ \mathrm{for \  haze} \qquad \ (1\ km\leq \tau \leq 6\ km),\\
             0 \qquad\qquad\quad\ \  \mathrm{for \  fog} \qquad\ \   (\tau \leq 0.5\ km). &
             \end{array}
\right.
\end{equation}
These transmission parameters are listed in Table~\ref{pathloss}.

Fig.~\ref{transmittance} illustrates how $\eta_{bt}$ varies with the transmission radius $R$ under three different air quality and two different beam wavelengths. We can see that, $\eta_{bt}$ decays exponentially with the increment of $R$. Meanwhile, for the same beam wavelength, $\eta_{bt}$ has much steeper shape with the decrement of $\tau$. It means that $\eta_{bt}$ declines faster with the air quality declining, i.e., the visibility gets low.

In addition, in the clear air and the haze scenarios, given the same radius $R$, the beam power attenuates faster for the short beam wavelength. When $\theta$ takes 0 in the fog scenario, for 810 nm and 1550 nm, $\eta_{bt}$ has the same attenuation pattern.

\subsubsection{Beam-to-Electricity Conversion}\label{}
At the ARBC receiver, various factors like the input beam power $P_{br}$, the beam wavelength $\lambda$, and the PV-cell temperature $T$ affect the PV-panel power conversion. We simulate the beam-to-electricity conversion procedure with the standard solar cell Simulink model \cite{dlciot}. Table~\ref{simulink} lists all the simulation  parameters.

When the PV-cell temperature $T$ is 25$^\circ$C (298K), Fig.~\ref{8101550Pirradiance} illustrates the influences that $P_{br}$ and $\lambda$ have on the PV-panel output power $P_{o,pv}$ based on \eqref{Iopv}, \eqref{Vm} and \eqref{popv}.

From Fig.~\ref{8101550Pirradiance}, $P_{o,pv}$ goes up gradually to the peak and then declines sharply to the bottom. The PV-panel outputs more power with the increment of $P_{br}$. The dots in Fig.~\ref{8101550Pirradiance} are the maximum power points (MPPs) of the curves, which have been proved that uniquely exist \cite{uniquempp}. Similarly, the MPPs when $T$ takes 0$^\circ$C (273K) and 50$^\circ$C (323K) can be obtained.

\begin{figure}
	\centering
	\includegraphics[scale=0.6]{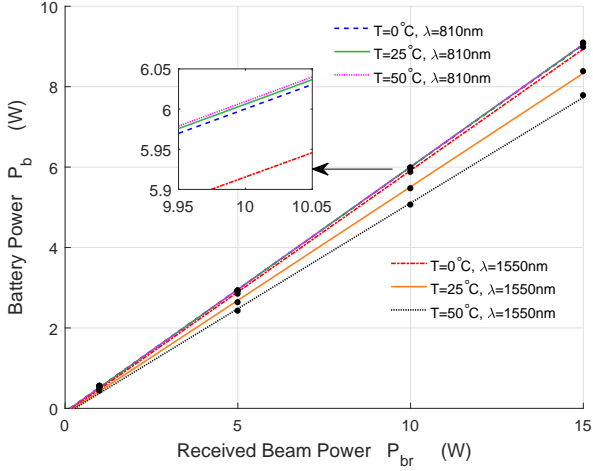}
	\caption{Battery Power vs. Received Beam Power}
	\label{8101550PbPlr}
\end{figure}

To obtain the maximum efficiency of the ARBC system, the PV-panel is expected to work at MPPs. Moreover, the maximum $P_{o,pv}$ should be the preferred battery charging power $P_{b}$. The dots in Fig.~\ref{8101550PbPlr} are the MPPs obtained under different PV-cell temperatures (0$^\circ$C, 25$^\circ$C and 50$^\circ$C) and beam wavelengths (810 nm and 1550 nm). From Fig.~\ref{8101550PbPlr}, the relationship between $P_{b}$ and $P_{br}$ can be obtained by the linear curve-fitting approximation as \cite{dlcvtc,dlciot}:
\begin{equation}\label{prpmf}
  P_{b} \approx a_2 P_{br} + b_2,
\end{equation}
where $a_2$ and $b_2$ are the linear curve fitting coefficients. To provide more details about how $P_{b}$ changes with $P_{br}$ under different temperature, values of $a_2$ and $b_2$ when the PV-cell temperature takes 0$^\circ$C, 5$^\circ$C, 10$^\circ$C, 15$^\circ$C, 20$^\circ$C, 25$^\circ$C, 30$^\circ$C, 35$^\circ$C, 40$^\circ$C, 45$^\circ$C, and 50$^\circ$C are listed in Table~\ref{a2b2}.

Fig.~\ref{8101550PbPlr} depicts the linear relationship between $P_{b}$ and $P_{br}$ for 810 nm and 1550 nm, respectively. The linear fitting lines match the measured $P_{b}$ and $P_{br}$, which are marked by the dots, very well. With the increment of $T$,  for same $P_{br}$, the value of $P_{b}$ diminishes. On the other hand, the 1550 nm system is more temperature-dependent than the 810 nm system.

\subsubsection{End-to-End Transmission}\label{}
Efficiency of battery charging, DC-DC conversion and feedback affect the end-to-end power transmission efficiency. We can assume there is almost no energy loss if the battery is charged with preferred power values. We also assume that the feedback system and the DC-DC converter cause almost no energy loss \cite{dcenergyloss}. Thus, the ARBC end-to-end power transmission mathematical model can be built based on the numerical models.

Based on \eqref{etalt}, \eqref{plps} and \eqref{prpmf}, given the beam transmission efficiency $\eta_{bt}$, the battery charging power $P_{b}$ changes depending on the supplied power $P_{s}$ as:
\begin{equation}\label{gspspm}
\begin{aligned}
  P_{b} &\approx a_2 \eta_{bt} P_{bt}+ b_2 \\
 & = a_1 a_2 \eta_{bt} P_{s} + (a_2 b_{1} \eta_{bt} + b_2).
\end{aligned}
\end{equation}

\begin{figure}
	\centering
	\includegraphics[scale=0.6]{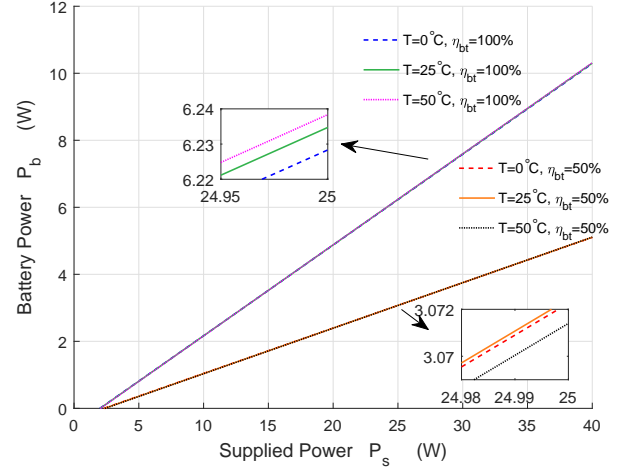}
	\caption{Battery Power vs. Supplied Power (810 nm)}
	\label{810pbps}
\end{figure}

\begin{figure}
	\centering
	\includegraphics[scale=0.6]{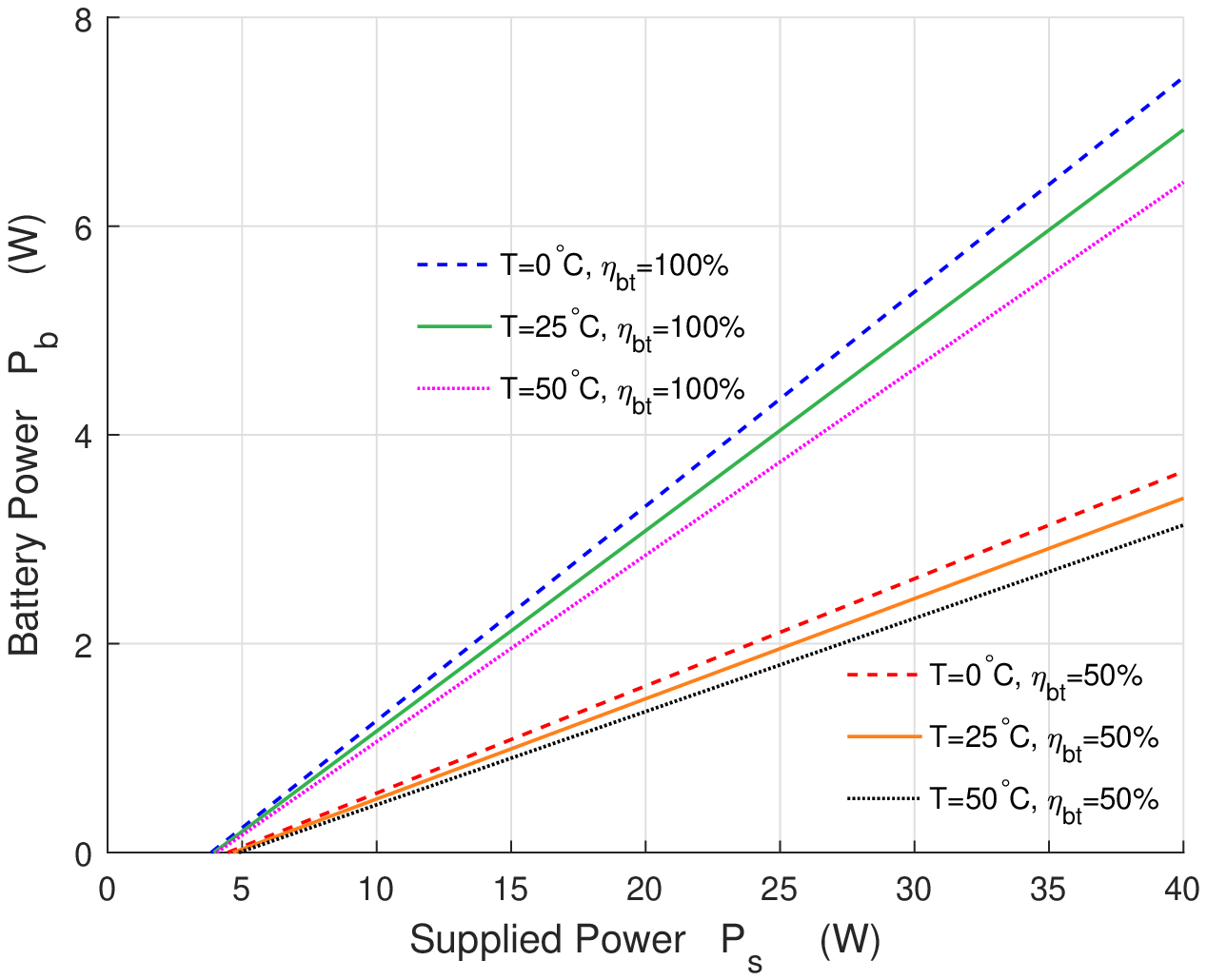}
	\caption{Battery Power vs. Supplied Power (1550 nm)}
	\label{1550pbps}
\end{figure}

Hence, $P_{b}$ depends on $P_{s}$ linearly. For the 810 nm system, Fig.~\ref{810pbps} depicts the linear relationship under three different $T$ when $\eta_{bt}$ takes 100\% and 50\%, respectively. While, Fig.~\ref{1550pbps} depicts the same relationships for the1550 nm system.

The linear relationship between $P_{b}$ and $P_{s}$ provides an intuitive and quantitive way to understand the end-to-end power transfer in the ARBC system.

\begin{algorithm}[b]
	\caption{Adaptive Resonant Beam Charging}\label{arbcalg}
	\begin{algorithmic}[1]
		
		\State \textbf{Begin}

        \State The power monitor gets the preferred charging power $P_b$, current $I_b$, voltage $V_b$ and cut-off time $t$     \label{checkcurrent}

        \While {$I_b$ $\geq$ 20 mA \textbf{and} $t$ $\le$ 3.6 h}
        \State $P_s \leftarrow P_b/(\eta_{eb}\eta_{bt}\eta_{pv})$  // The power monitor computes $P_s$, and sends it to the power controller
        \State $P_{bt} \leftarrow P_s\eta_{eb}$         // Transmitter beam power
        \State $P_{br} \leftarrow P_{bt}\eta_{bt}$   // Receiver beam power
        \State $P_{b} \leftarrow P_{br}\eta_{pv}$   // PV-panel output power
        \If {$I_{pv}$ $\not=$ $I_b$ \textbf{and} $V_{pv}$ $\not=$ $V_b$}   // DC-DC conversion
		\State $I_b \leftarrow I_{pv}$
		\State $V_b \rightarrow V_{pv} $
		\EndIf
        \State Charge the battery with $I_b$ and $V_b$
        \State The power monitor updates $P_b$, $I_b$ and $V_b$ according to the next battery state

		\EndWhile

        \State Stop Charging
		
		\State \textbf{End}

	\end{algorithmic}
\end{algorithm}

\begin{table*}[b]
\centering
\caption{Supplied Energy Consumption in Different Scenarios (Wh)}
 \begin{tabular}{C{1.1cm} C{2.6cm} C{4cm}  C{4cm}  C{4cm}}
\hline
 {} & \bfseries{Temperature $T$}  &0$^\circ$C&25$^\circ$C &50$^\circ$C \\
 {} & \bfseries{Distance $R$} & {0.1\ KM \qquad 0.5\ KM \qquad 1\ KM\qquad} & {0.1\ KM \qquad 0.5\ KM \qquad 1\ KM\qquad} & {0.1\ KM \qquad 0.5\ KM \qquad 1\ KM\qquad} \\
\hline
 {} & \bfseries{ \qquad\qquad\quad Clear air} & {\ \ 64.20 \qquad\  69.91 \qquad\quad  77.86\quad} & {\ \ 64.15 \qquad\ 69.85 \qquad\quad 77.79\quad}& {\ \ 64.15 \qquad\ 69.86 \qquad \quad 77.80}\\
 {\bfseries{810 nm}} & \bfseries{ RBC \qquad\ Haze} & {\ \ 68.46 \qquad\  97.02 \qquad\quad 152.01\quad} & {\ \ 68.39 \qquad\  96.93 \qquad\quad 151.86\quad}& {\ \ 68.40 \qquad\  96.94 \qquad\quad 151.88\quad}\\
 {} & \bfseries{ \qquad\qquad\quad Fog} & {\ 156.23 \ \  7.59$\times10^{3}$ \ \ 1.0115$\times10^{6}$} & {\ 156.08 \  7.579$\times10^{3}$ \  1.0105$\times10^{6}$}& {\ 156.09 \  7.58$\times10^{3}$ \ \ 1.0106$\times10^{6}$}\\
\midrule
 {} & \bfseries{ \qquad\qquad\quad Clear air} & {\ \ 29.97 \qquad\  32.28 \qquad\quad 35.49\quad} & {\ \ 29.99 \qquad\  32.31 \qquad\quad 35.52\quad}& {\ \ 30.05 \qquad\  32.36 \qquad\quad 35.58\quad}\\
 {\bfseries{810 nm}} & \bfseries{ARBC \qquad Haze} & {\ \ 31.69 \qquad\  43.23 \qquad\quad 65.45\quad} & {\ \ 31.72 \qquad\  43.27 \qquad\quad 65.51\quad}& {\ \ 31.77 \qquad\  43.35 \qquad\quad 65.63\quad} \\
 {} & \bfseries{ \qquad\qquad\quad Fog} & {\ 67.15 \ \ 3.069$\times10^{3}$ \ \ 4.087$\times10^{5}$} & {67.22 \  3.073$\times10^{3}$ \ \ 4.091$\times10^{5}$}& {\ 67.34 \quad 3.079$\times10^{3}$ \ 4.099$\times10^{5}$}\\
\midrule
 {} & \bfseries{ \qquad\qquad\quad Clear air} & {\  65.15 \qquad\  70.96 \qquad\quad 79.03\quad} & {\ \ 69.40 \qquad\  75.63 \qquad\quad 84.30\quad}& {\ \ 74.24 \qquad\  80.94 \qquad\quad 90.28\quad}\\
 {\bfseries{1550 nm}} & \bfseries{ RBC \qquad Haze} & {\ \ 69.47 \qquad\  98.51 \qquad\quad 154.40\quad} & {\ \ 74.04 \qquad\  105.19 \qquad 165.15}& {\ \ 79.23 \qquad\  112.78 \qquad 177.37} \\
 {} & \bfseries{ \qquad\qquad\quad Fog} & {\ 158.69 \ \ 7.71$\times10^{3}$ \ \ \ 1.03$\times10^{6}$} & {169.75 \  8.27$\times10^{3}$ \ \ 1.10$\times10^{6}$}& {\ 182.33 \ \ \  8.91$\times10^{3}$ \ \ 1.19$\times10^{6}$}\\
\midrule
 {} & \bfseries{ \qquad\qquad\quad Clear air} & {\ \ 30.71 \qquad\  33.09 \qquad\quad 36.41\quad} & {\ \ 32.55 \qquad\  35.11 \qquad\quad 38.68\quad}& {\ \ 34.63 \qquad\  37.40 \qquad\quad 41.26\quad}\\
 {\bfseries{1550 nm}} & \bfseries{ARBC \qquad Haze} & {\ \ 32.48 \qquad\  44.39 \qquad\quad 67.31\quad} & {\ \ 34.46 \qquad\  47.28 \qquad\quad 71.96\quad}& {\ \ 36.69 \qquad\  50.55 \qquad\quad 77.22\quad} \\
 {} & \bfseries{ \qquad\qquad\quad Fog} & {\ \ 69.07 \ \ 3.17$\times10^{3}$ \ \ \ 4.22$\times10^{5}$} & {\ \ 73.85 \ \  3.41$\times10^{3}$ \ \ \ 4.54$\times10^{5}$}& {\ \ 79.27 \ \  3.68$\times10^{3}$ \ \ \ 4.91$\times10^{5}$}\\
\hline
\label{consumedenergyRadius}
\end{tabular}
\end{table*}

\subsection{Control Algorithm}\label{}
Based on the mathematical modeling, given the battery preferred charging current, voltage, and thus power, the supplied power can be calculated out with reference to the end-to-end transmission efficiency. To demonstrate the charging operation of the ARBC system, we give the charging algorithm in Algorithm 1 as:

1) The power controller gets the battery preferred charging power $P_b$, current $I_b$, voltage $V_b$ or cut-off time $t$ from the power monitor.

2) If $I_b$ is lower than 20 mA or $t$ equals to 3.6 h, the charging procedure ends. Otherwise, the power controller informs the power supply to generate $P_s$ from $P_b$ with reference to the end-to-end transmission efficiency.

3) $P_s$ effects on the gain medium, and the resonant beam can be stimulated out.

4) The PV-panel can convert the beam power to the electrical power $P_{pv}$, while the output current and voltage are $I_{pv}$ and $V_{pv}$.

5) The DC-DC converter converts $I_{pv}$ and $V_{pv}$ to $I_b$ and $V_b$. Thus, the battery can be charged with preferred values.

6) The power monitor updates $I_b$ and $V_b$ according to the next battery state, and sends them to the power controller. Then, turn to 2).

Repeating these steps, the battery can be charged with the battery preferred charging values dynamically during the whole ARBC procedure.

\section{Performance Evaluation}\label{}
Based on the ARBC system design, the batteries accessed to the ARBC system can be charged with their preferred values. Therefore, the system performance can be evaluated. In terms of battery charging and power supply, the advantages of the ARBC system over the RBC system will be validated by the performance comparisons in this section. The numerical evaluations are implemented in MATLAB and Simulink.

\subsection{Battery Charging Performance}\label{}

\begin{figure}
	\centering
	\includegraphics[scale=0.6]{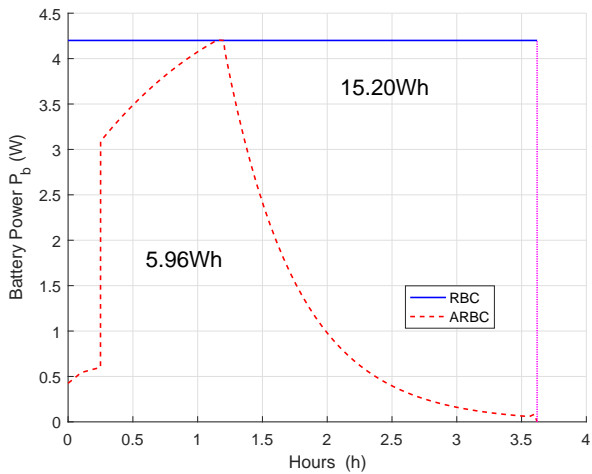}
	\caption{Battery Power in ARBC and RBC}
	\label{batterypower}
\end{figure}

In traditional wireless charging systems, including the RBC system, charging stages can not be tracked without adaptive function. Hence, batteries are charged with fixed current or voltage. We take 4.2 W (constant current 1 A and constant voltage 4.2 V) for Li-ion battery charging power. The solid-line in Fig.~\ref{batterypower} shows the constant battery charging power. It is an horizontal line without any changes during the RBC procedure.

\begin{figure}
	\centering
	\includegraphics[scale=0.6]{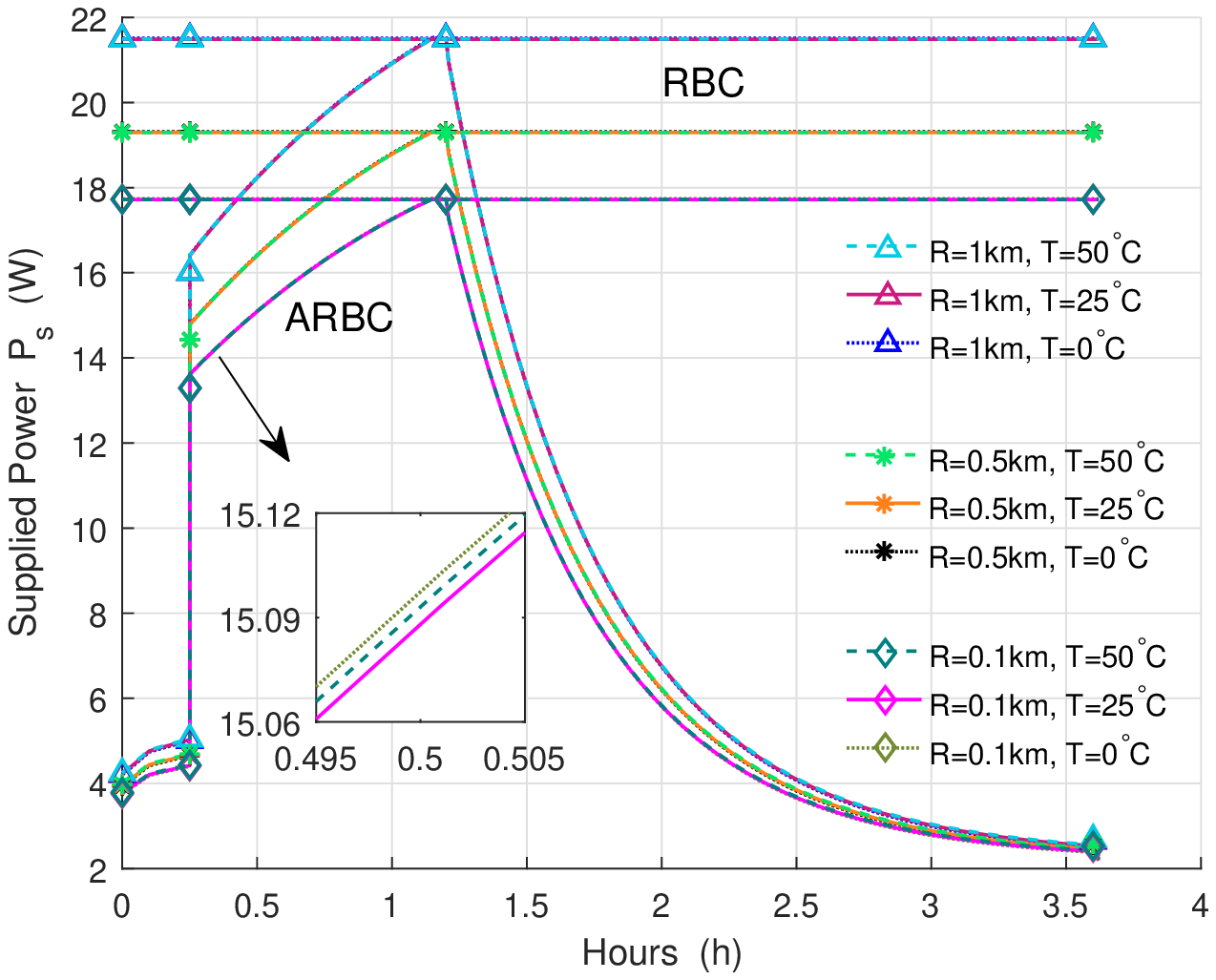}
	\caption{Supplied Power with Different Radius and Temperature for 810 nm (Clear Air)}
	\label{pst810clear}
%
%
	\centering
	\includegraphics[scale=0.6]{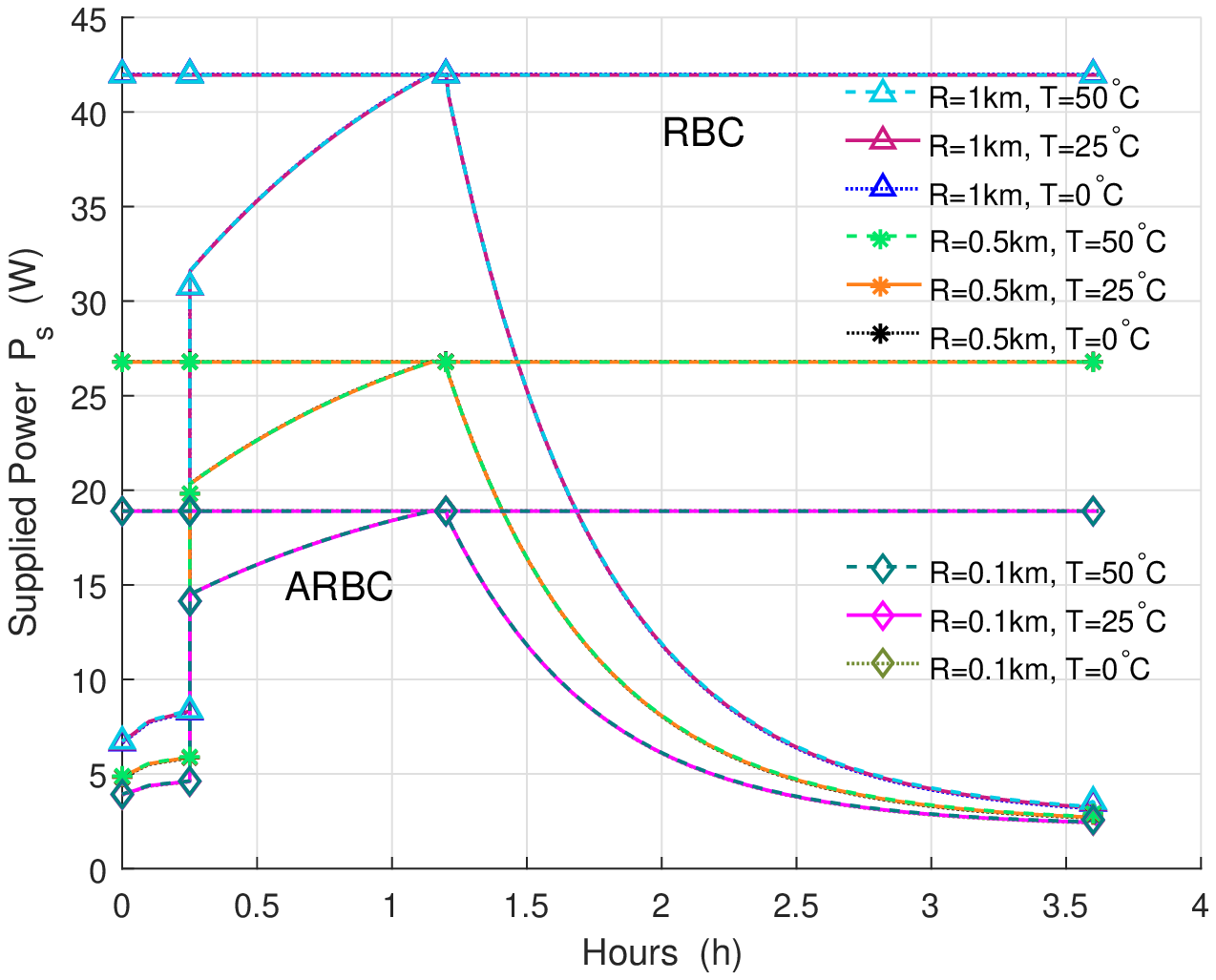}
	\caption{Supplied Power with Different Radius and Temperature for 810 nm (Haze)}
	\label{pst810haze}
%
	\centering
	\includegraphics[scale=0.6]{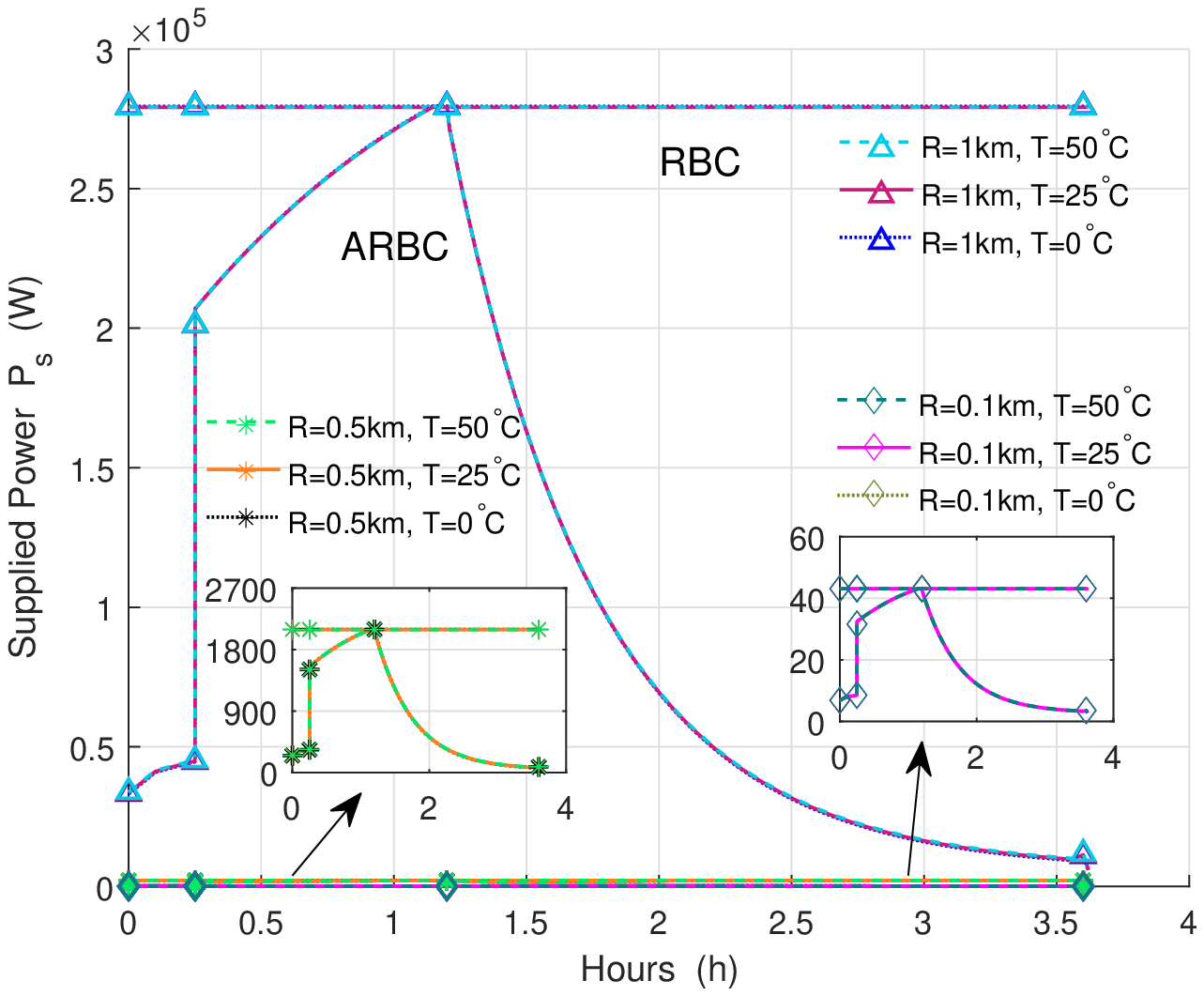}
	\caption{Supplied Power with Different Radius and Temperature for 810 nm (Fog)}
	\label{pst810fog}
\end{figure}

\begin{figure}
	\centering
	\includegraphics[scale=0.6]{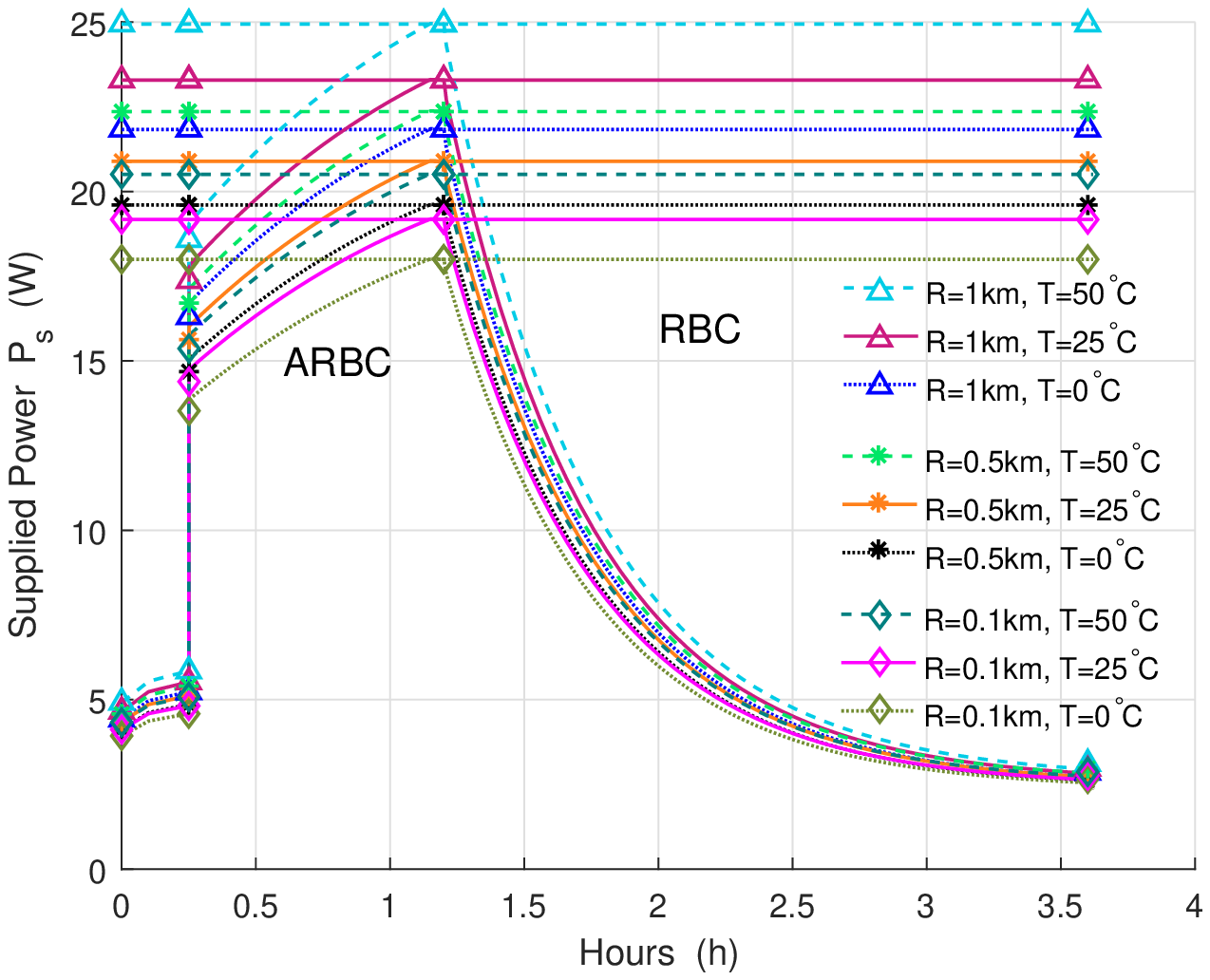}
	\caption{Supplied Power with Different Radius and Temperature for 1550 nm (Clear Air)}
	\label{pst1550clear}
%
	\centering
    \includegraphics[scale=0.6]{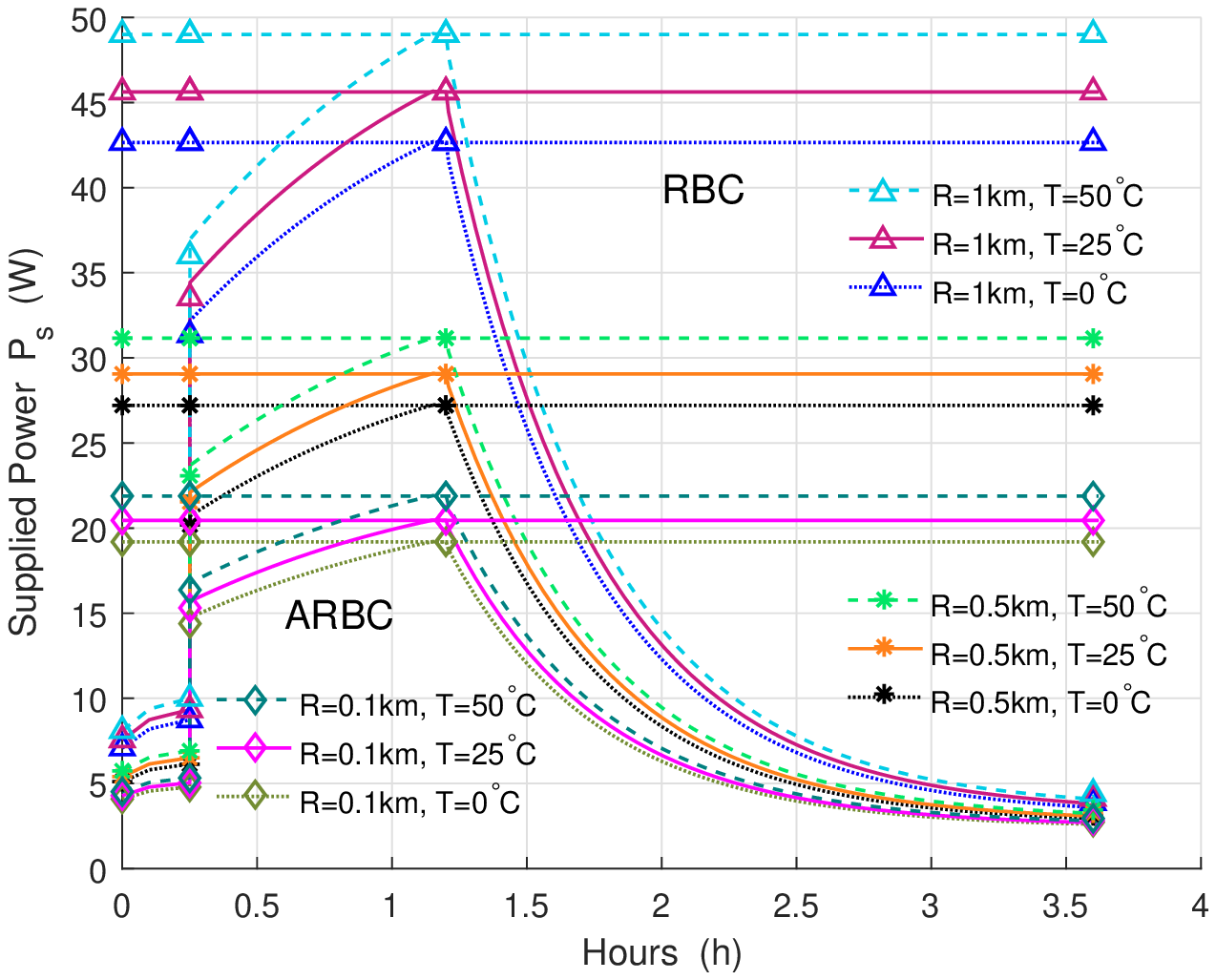}
	\caption{Supplied Power with Different Radius and Temperature for 1550 nm (Haze)}
	\label{pst1550haze}
   \centering
    \includegraphics[scale=0.6]{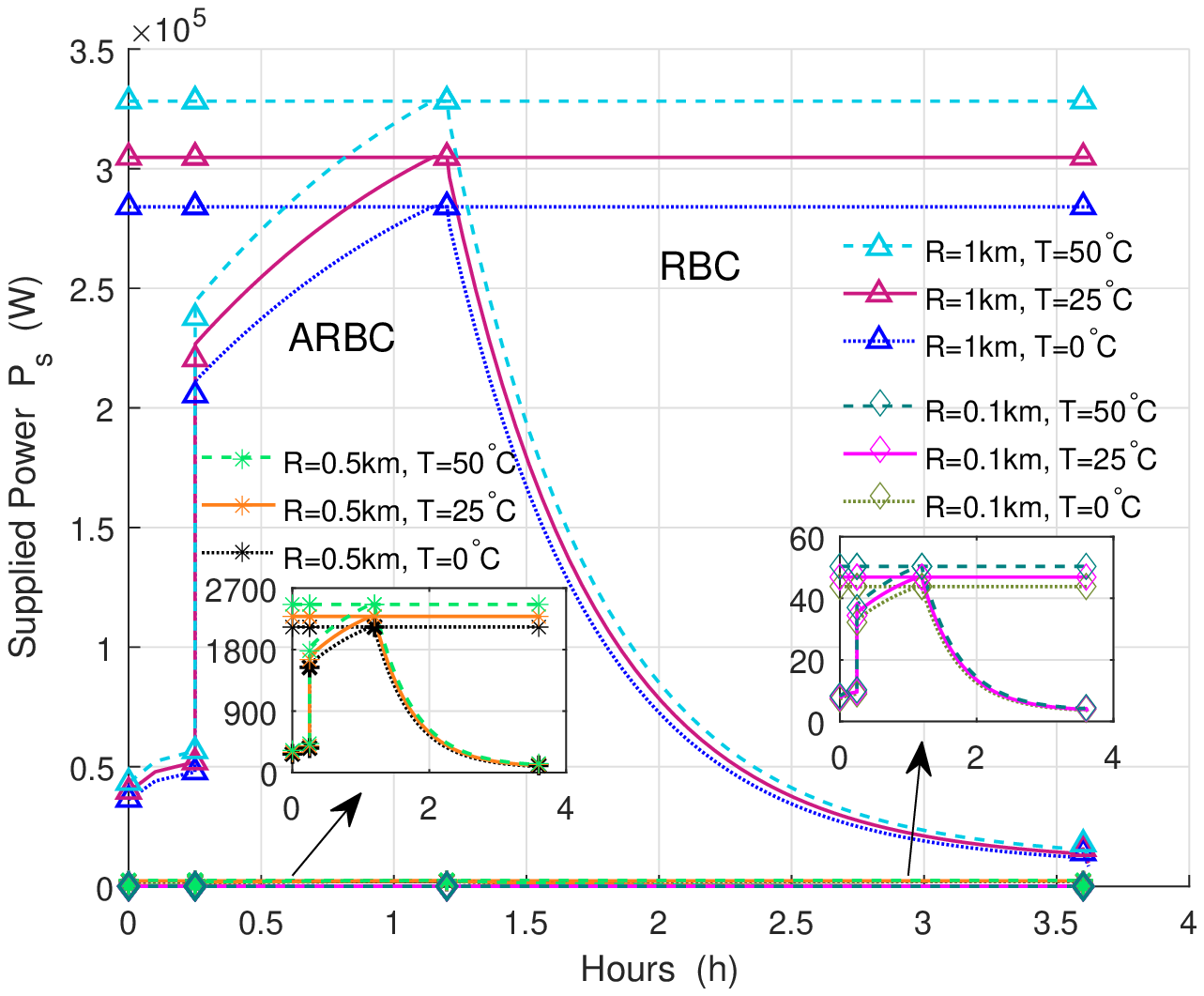}
	\caption{Supplied Power with Different Radius and Temperature for 1550 nm (Fog)}
	\label{pst1550fog}
\end{figure}

\begin{figure}
	\centering
	\includegraphics[scale=0.55]{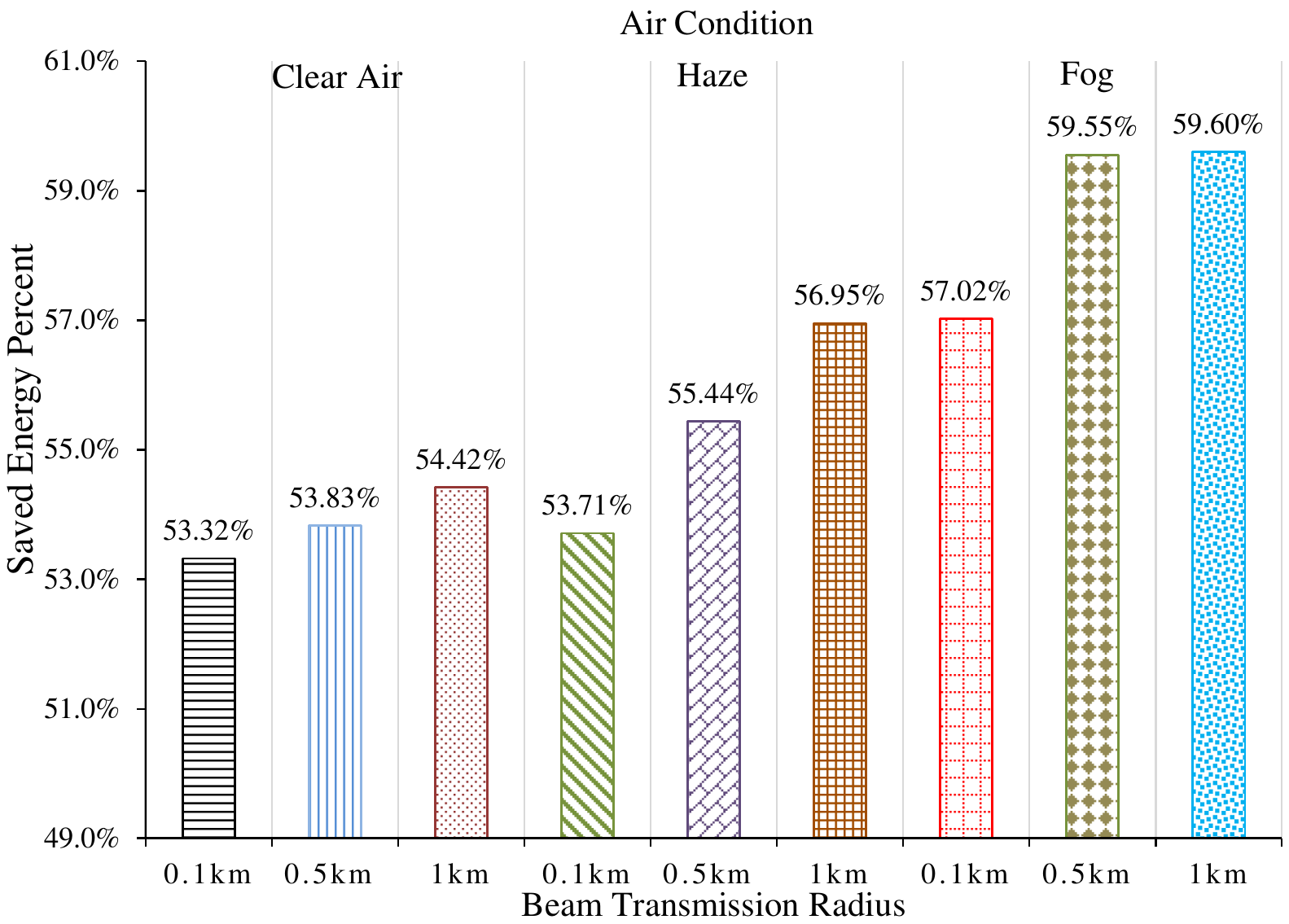}
	\caption{The Percent of Saved Supplied Energy under 0$^\circ$C ($\lambda$=810 nm)}
	\label{savedenergyRadius810273}
%
	\centering
	\includegraphics[scale=0.55]{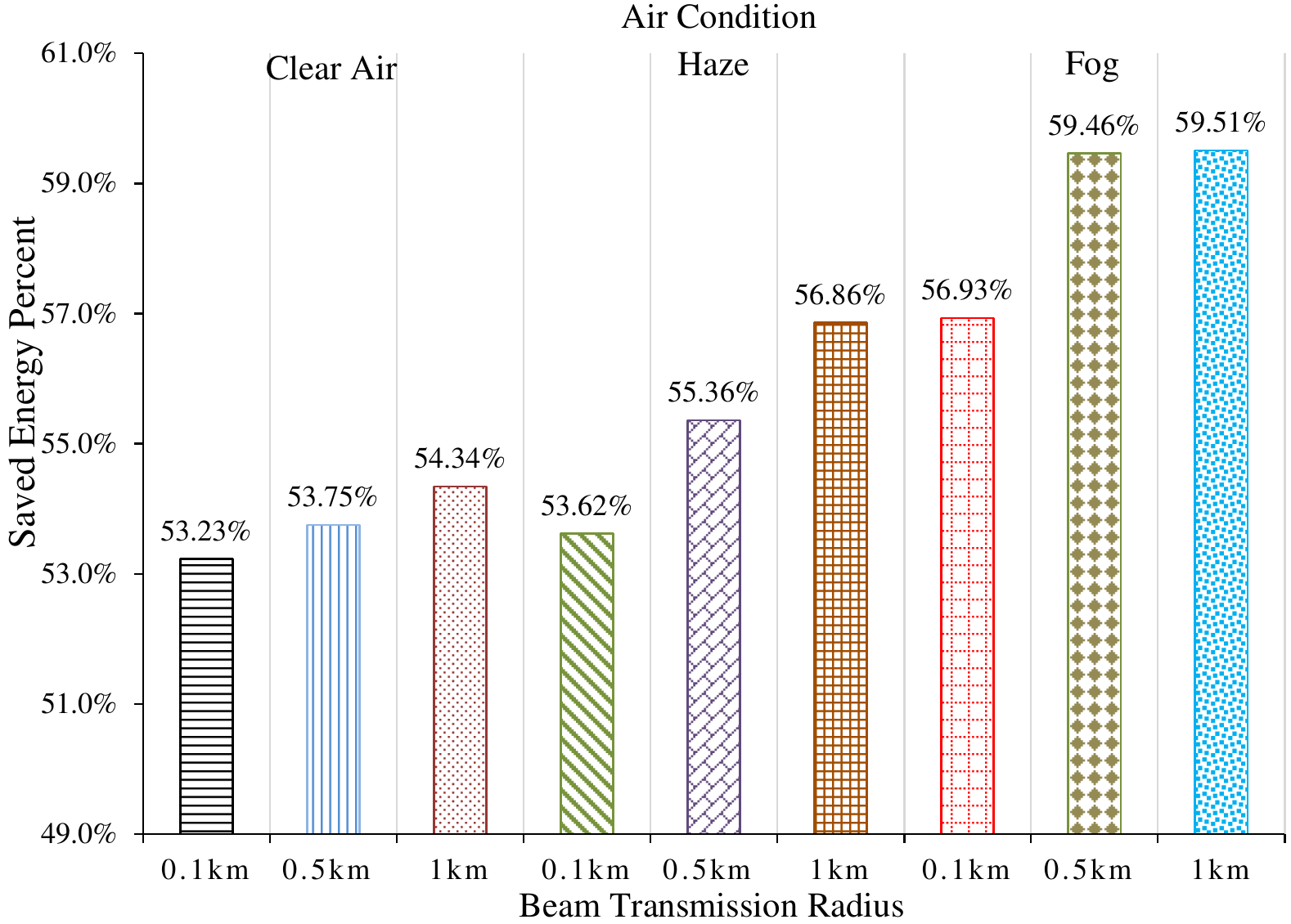}
	\caption{The Percent of Saved Supplied Energy under 25$^\circ$C ($\lambda$=810 nm)}
	\label{savedenergyRadius810298}
%
	\centering
	\includegraphics[scale=0.55]{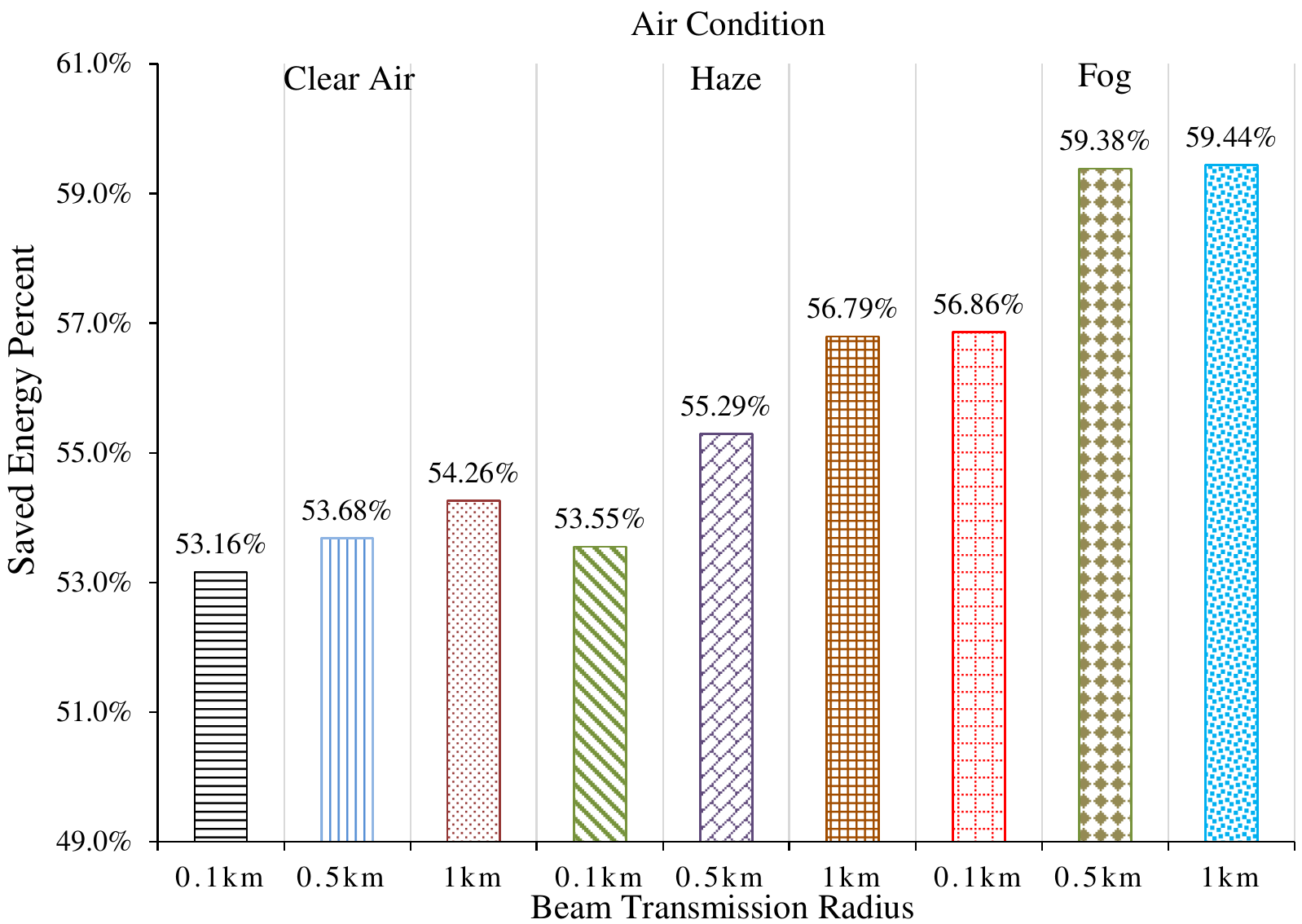}
	\caption{The Percent of Saved Supplied Energy under 50$^\circ$C ($\lambda$=810 nm)}
	\label{savedenergyRadius810323}
\end{figure}

\begin{figure}
	\centering
	\includegraphics[scale=0.55]{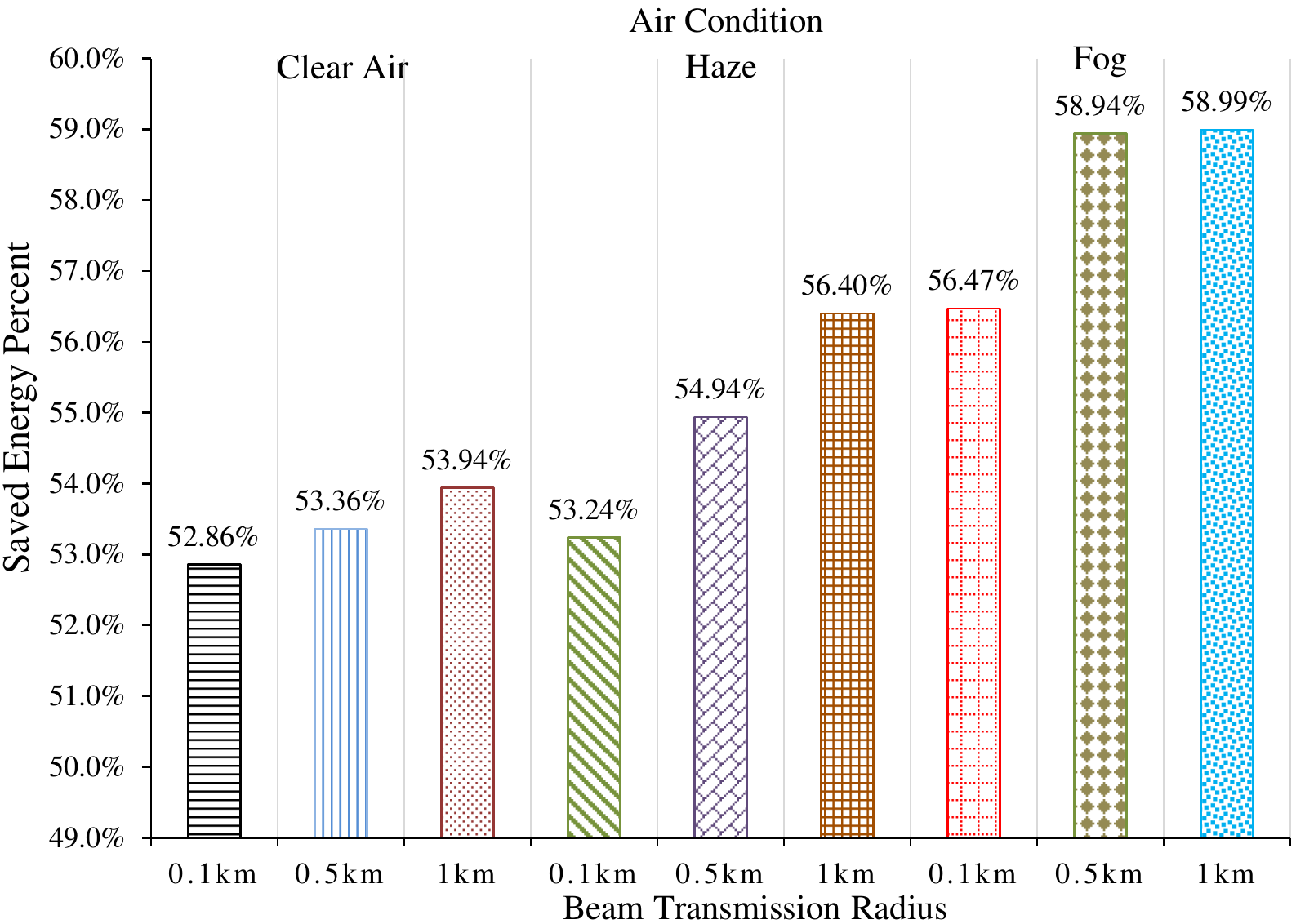}
	\caption{The Percent of Saved Supplied Energy under 0$^\circ$C ($\lambda$=1550 nm)}
	\label{savedenergyRadius1550273}
%
	\centering
	\includegraphics[scale=0.55]{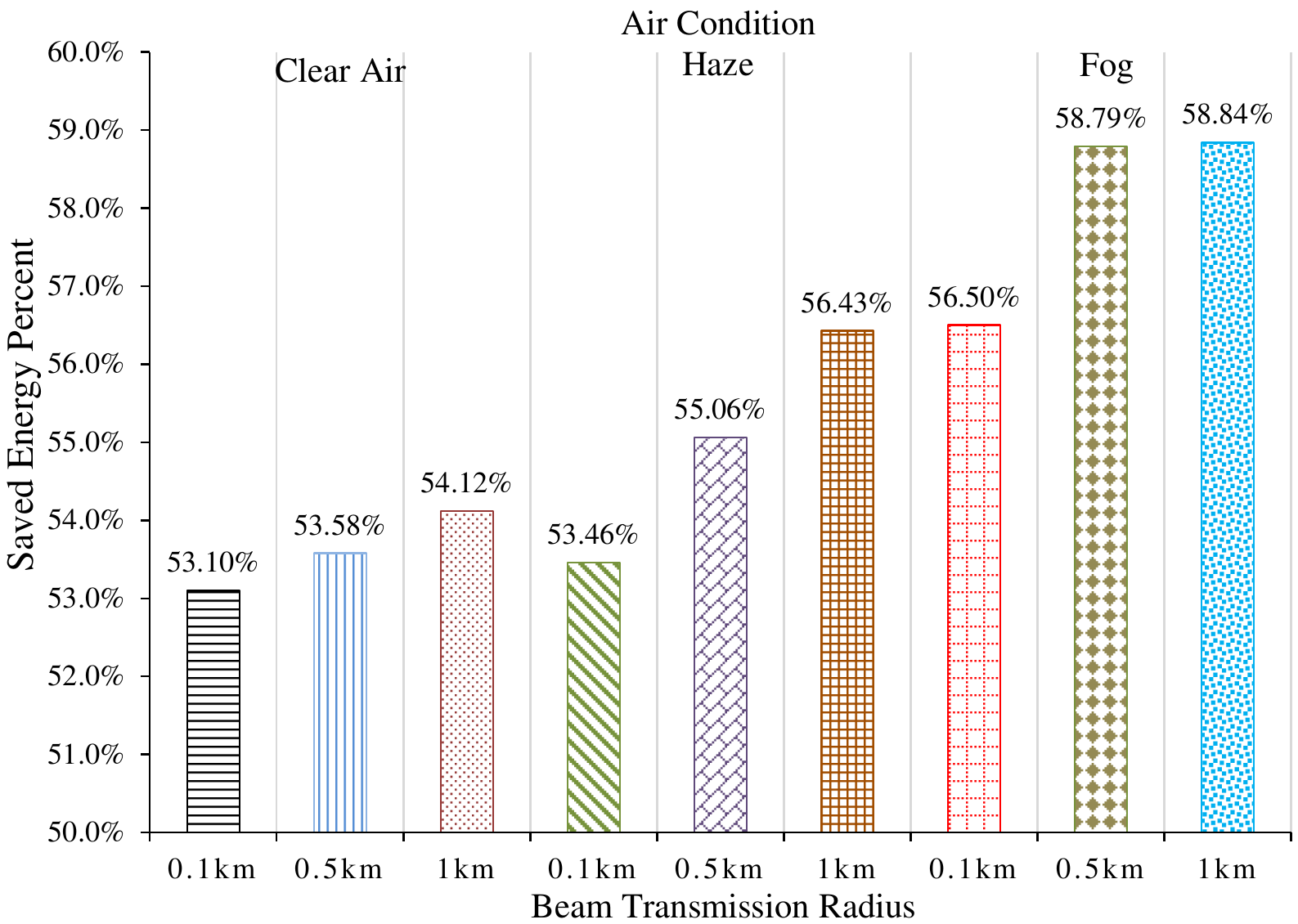}
	\caption{The Percent of Saved Supplied Energy under 25$^\circ$C ($\lambda$=1550 nm)}
	\label{savedenergyRadius1550298}
%
	\centering
	\includegraphics[scale=0.55]{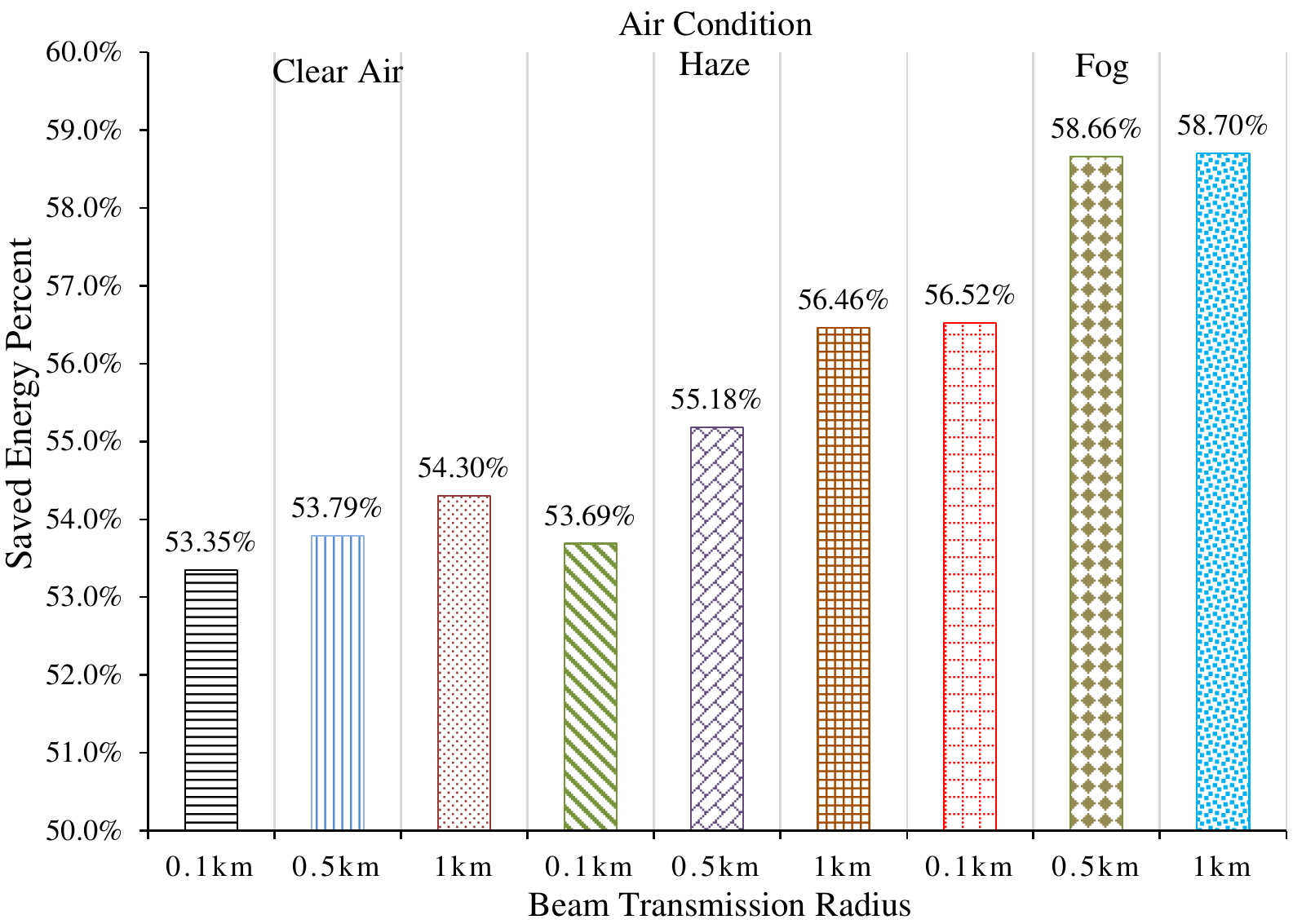}
	\caption{The Percent of Saved Supplied Energy under 50$^\circ$C ($\lambda$=1550 nm)}
	\label{savedenergyRadius1550323}
\end{figure}

However, batteries can be charged dynamically in the ARBC system. According to the Li-ion battery charging profile in Fig.~\ref{li-ionchargezero}, the preferred battery charging power can be obtained according to the preferred charging current and voltage. The dash-line in Fig.~\ref{batterypower} shows the preferred battery charging power changes during the ARBC procedure.

From Fig.~\ref{batterypower}, we can recognize the big power consumption gap between the RBC system and the ARBC system. The areas under the battery charging power in Fig.~\ref{batterypower} stand for the consumed battery charging energy. The RBC procedure consumed energy is 15.20 Wh, while the ARBC procedure only consumes 5.96 Wh. That is to say, during the whole charging procedure, the ARBC system could save 9.24 Wh energy. In summary, 61\% battery charging energy can be saved by the ARBC system compared with the RBC system.

\subsection{Power Supply Performance}\label{}
For different transmission radius $R$, air quality, beam wavelength $\lambda$, and PV-cell temperature $T$, the supplied power $P_s$ takes different values in the ARBC system. That is, $P_s$ depends on $R$, $T$, $\lambda$ and air quality. For $\lambda$=810 nm, Figs.~\ref{pst810clear}, \ref{pst810haze} and \ref{pst810fog} show how $P_s$ changes with different radius $R$ under different air quality (clear air, haze and fog) and different temperature $T$ (0$^\circ$C, 25$^\circ$C and 50$^\circ$C), respectively. Figs.~\ref{pst1550clear}, \ref{pst1550haze}, and \ref{pst1550fog} illustrate $P_s$ in the same scenarios for $\lambda$=1550 nm, respectively.

Form Figs.~\ref{pst810clear}-\ref{pst1550fog}, when $R$ increases, $\eta_{bt}$ becomes small. Thus, $P_s$ needs to be enhanced to compensate the power attenuation. Since the PV-panel takes lower conversion efficiency as $T$ goes up, $P_s$ and $T$ show a negative correlation. Moreover, with the same $R$, $T$ and $\lambda$, $P_s$ increases when the air visibility decreases. This is due to the attenuation of $\eta_{bt}$ becomes higher with the air quality getting worse. To obtain the preferred battery charging power, $P_s$ increases as $\eta_{bt}$ attenuates.

Then, the corresponding consumed supplied energy during the charging procedure can be obtained as in Table~\ref{consumedenergyRadius}. It can be observed that, for the same beam wavelength, the consumed supplied power keeps the upward trend as $T$ increases within certain radius under certain air quality. The ARBC system consumes much less supplied energy in all the listed scenarios for both 810 nm and 1550 nm compared with the RBC system.

Thereafter, for the same charging procedure, the supplied energy saved by the ARBC system compared with the RBC system can then be obtained. Figs.~\ref{savedenergyRadius810273}, \ref{savedenergyRadius810298} and \ref{savedenergyRadius810323} show the saved supplied energy percentage for 810 nm under 0$^\circ$C, 25$^\circ$C and 50$^\circ$C, respectively. Figs.~\ref{savedenergyRadius1550273}, \ref{savedenergyRadius1550298} and \ref{savedenergyRadius1550323} show the saved supplied energy percentage in the same scenarios for 1550 nm.

From Figs.~\ref{savedenergyRadius810273}-\ref{savedenergyRadius1550323}, under the same air quality, the percentage of saved energy goes up with $R$ increasing under the same $T$. With the same $R$ and $T$, the percentage of saved energy goes up as the air quality declines. Since whether $R$ increases or air quality declines, $\eta_{bt}$ decreases, and thus $P_s$ needs to be increased. The increment of $P_s$ leads to thermal effect and energy loss. Therefore, to obtain same battery charging power, the smaller $P_s$ is preferred by ARBC compared with RBC. Hence, less energy will loss in the ARBC system, and more energy will be saved. This validates that the necessity of adopting the ARBC system increases as the preferred supplied power $P_s$ rising. Moreover, in the above scenarios, 53\%-60\% supplied energy can be saved by the ARBC system no matter for the 810 nm or the 1550 nm system.

\subsection{Summary}\label{}
The ARBC mechanism introduced in Section \ref{Section3} is numerically evaluated. Based on the end-to-end power conversion analysis, we quantitatively demonstrate the variation of the supplied power during battery charging period procedure. The supplied power depends on the PV-cell temperature, beam transmission efficiency, and beam wavelength. Furthermore, we obtain the battery charging energy and the supplied energy saved by the ARBC system compared with the RBC system in different scenarios.

The observations from the above analysis include:
\begin{itemize}
  \item With the increment of either the transmission radius $R$ or the PV-cell temperature $T$, the preferred supplied power $P_s$ takes an upward trend in the ARBC system.
  \item The supplied power $P_s$ shows a negative correlation with the air quality. That is to say, with the improvement of the air quality, i.e., the enhancement of the visibility, less supplied power is required to get the certain battery charging power.
  \item For the same air quality and transmission radius $R$, the consumed supplied energy goes up with the PV-cell temperature $T$ increasing for the RBC system and the ARBC system.
  \item For the same air quality, the percentage of the ARBC system saved energy goes up as the transmission radius $R$  increases.
  \item For the same transmission radius $R$, the ARBC system saves more energy with the air quality  going down.
  \item For battery charging, 61\% energy can be saved by the ARBC system compared with the RBC system.
  \item For the power supply, 53\%-60\% supplied energy is saved by the ARBC system compared with the RBC system.
\end{itemize}


\section{Conclusions}\label{Conclusions}

An adaptive resonant beam charging (ARBC) system is introduced in this paper based on the resonant beam charging system (RBC). The system design and numerical analysis of the ARBC system are presented to optimize battery charging performance in terms of battery charging profile. Given the supplied power, the battery charging power is influenced by various factors, including beam wavelength, beam transmission efficiency and PV-cell temperature etc.. Numerical analysis illustrates that 61\% battery charging energy and 53\%-60\% supplied energy can be saved by the ARBC system compared with the RBC system.

Several issues in this area are worth of further research:
\begin{itemize}
  \item The analysis in this paper is under ideal assumptions. For example, the feedback and measurement errors are inevitable in the practical system, which should be considered in the future.
  \item For different beam wavelengths, the saved energy varies as the PV-cell temperature changing
  \item Different battery types have different battery charging profiles. Therefore, studying their impacts on ARBC is one of the future research areas.
\end{itemize}


\section{Acknowledgment}\label{acknowledge}

The authors would like to thank the editors and the anonymous reviewers. At the same time, we would like to thank colleagues in our laboratory. Thank Hao Deng and Mingqing Liu for their valuable suggestions, and thank Aozhou Wu for polishing the figures.

{
\normalsize
\bibliographystyle{IEEEtran}
\bibliographystyle{unsrt}
\bibliography{references}

\begin{thebibliography}{10}
\providecommand{\url}[1]{#1}
\csname url@samestyle\endcsname
\providecommand{\newblock}{\relax}
\providecommand{\bibinfo}[2]{#2}
\providecommand{\BIBentrySTDinterwordspacing}{\spaceskip=0pt\relax}
\providecommand{\BIBentryALTinterwordstretchfactor}{4}
\providecommand{\BIBentryALTinterwordspacing}{\spaceskip=\fontdimen2\font plus
\BIBentryALTinterwordstretchfactor\fontdimen3\font minus
  \fontdimen4\font\relax}
\providecommand{\BIBforeignlanguage}[2]{{%
\expandafter\ifx\csname l@#1\endcsname\relax
\typeout{** WARNING: IEEEtran.bst: No hyphenation pattern has been}%
\typeout{** loaded for the language `#1'. Using the pattern for}%
\typeout{** the default language instead.}%
\else
\language=\csname l@#1\endcsname
\fi
#2}}
\providecommand{\BIBdecl}{\relax}
\BIBdecl

\bibitem{ding2014cognitive}
Q.~Wu, G.~Ding, Y.~Xu, Z.~S.~Feng, J.~Du, Wang, and K.~Long, ``Cognitive
  internet of things: a new paradigm beyond connection,'' \emph{IEEE Internet
  of Things Journal}, vol.~1, no.~2, pp. 129--143, Apr. 2014.

\bibitem{multimediasignals}
S.~Stankovi\'{c}, I.~Orovi\'{c}, and E.~Sejdi\'{c}, \emph{Multimedia signals
  and systems}, 2nd~ed.\hskip 1em plus 0.5em minus 0.4em\relax New York:
  Springer Cham, 2016.

\bibitem{yu2014survey}
S.~Peng, S.~Yu, and A.~Yang, ``Smartphone malware and its propagation modeling:
  {A} survey,'' \emph{IEEE Communications Surveys \& Tutorials}, vol.~16,
  no.~2, pp. 925--941, 2014.

\bibitem{wirelesspathent}
J.~M. Fernandez and J.~A. Borras, ``Contactless battery charger with wireless
  control link,'' USA Patent 5\,668\,842, Feb., 2001.

\bibitem{wirelesspaper1}
A.~Esser and H.-C. Skudelny, ``A new approach to power supplies for robots,''
  \emph{IEEE Transactions on Industry Applications}, vol.~27, no.~5, pp.
  872--875, Oct. 1991.

\bibitem{wirelesspaper2}
J.~Hirai, T.~W. Kim, and A.~Kawamur, ``Wireless transmission of power and
  information for cableless linear motor drive,'' \emph{IEEE Transactions on
  Power Electron}, vol.~15, no.~1, pp. 21--27, Jan. 2000.

\bibitem{inductive}
S.~E. Sarma, S.~A. Weis, and D.~W. Engels, ``Rfid systems and security and
  privacy implications,'' in \emph{International Workshop on Cryptographic
  Hardware and Embedded Systems}.\hskip 1em plus 0.5em minus 0.4em\relax
  Springer, Feb. 2003, pp. 454--469.

\bibitem{magnetic}
A.~Kurs, A.~Karalis, R.~Moffatt, J.~D. Joannopoulos, P.~Fisher, and
  M.~Solja{\v{c}}i{\'c}, ``Wireless power transfer via strongly coupled
  magnetic resonances,'' \emph{science}, vol. 317, no. 5834, pp. 83--86, Jul.
  2007.

\bibitem{radio}
W.~G. Bigelow, J.~C. Callaghan, and J.~A. Hopps, ``General hypothermia for
  experimental intracardiac surgery: {T}he use of electrophrenic respirations,
  an artificial pacemaker for cardiac standstill, and radio-frequency rewarming
  in general hypothermia,'' \emph{Annals of surgery}, vol. 132, no.~3, p. 531,
  Sept. 1950.

\bibitem{laser}
D.~U. Bartsch, I.~K. Muftuoglu, and W.~R. Freeman, ``Laser pointers revisited
  editorial,'' \emph{Retina}, vol.~36, no.~9, pp. 1611--1613, Sept. 2016.

\bibitem{wicharge}
R.~Della-Pergola, O.~Alpert, O.~Nahmias, and V.~Vaisleib, ``Spatiallity
  distributed laser resonator,'' Israel Patent 20\,140\,126\,603, May, 2014.

\bibitem{yu2017internet}
B.~Feng, H.~Zhang, H.~Zhou, and S.~Yu, ``Locator/{I}dentifier split networking:
  {A} promising future internet architecture,'' \emph{IEEE Communications
  Surveys \& Tutorials}, vol.~19, no.~4, pp. 2927--2948, Fourthquarter 2017.

\bibitem{liu2016dlc}
Q.~Liu, J.~Wu, P.~Xia, S.~Zhao, W.~Chen, Y.~Yang, and L.~Hanzo, ``Charging
  unplugged: Will distributed laser charging for mobile wireless power transfer
  work?'' \emph{IEEE Vehcular Technology Magzine}, vol.~11, no.~4, pp. 36--45,
  Nov. 2016.

\bibitem{dlcvtc}
Q.~Zhang, X.~Shi, Q.~Liu, J.~Wu, P.~Xia, and Y.~Liao, ``Adaptive distributed
  laser charging for efficient wireless power transfer,'' in \emph{Vehicular
  Technology Conference (VTC-Fall), 2017 IEEE 86th}, Sept. 2017, pp. 1--5.

\bibitem{dlciot}
Q.~Zhang, W.~Fang, Q.~Liu, J.~Wu, P.~Xia, and L.~Yang, ``Distributed laser
  charging: {A} wireless power transfer approach (in press),'' \emph{IEEE
  Internet of Things Journal}, Jun. 2018,
  \mbox{doi}:\url{10.1109/JIOT.2018.2851070}.

\bibitem{fafc}
W.~Fang, Q.~Zhang, Q.~Liu, J.~Wu, and P.~Xia, ``Fair scheduling in resonant
  beam charging for {I}o{T} devices (in press),'' \emph{IEEE Internet of Things
  Journal}, Jul. 2018, \mbox{doi}:\url{10.1109/JIOT.2018.2853546}.

\bibitem{goodenough2013li}
J.~B. Goodenough and K.~Park, ``The {L}i-ion rechargeable battery: {A}
  perspective,'' \emph{Journal of the American Chemical Society}, vol. 135,
  no.~4, pp. 1167--1176, Jan. 2013.

\bibitem{shen2012charging}
W.~Shen, T.~T. Vo, and A.~Kapoor, ``Charging algorithms of lithium-ion
  batteries: An overview,'' in \emph{2012 7th IEEE Conference on Industrial
  Electronics and Applications (ICIEA)}, Jul. 2012, pp. 1567--1572.

\bibitem{laserdiodes}
B.~V.~Zeghbroeck, \emph{Principles of semiconductor devices}, 1st~ed.\hskip 1em
  plus 0.5em minus 0.4em\relax University of Colorado, 2004.

\bibitem{attenuation}
S.~A. Salman, J.~M. Khalel, and W.~H. Abas, ``Attenuation of infrared laser
  beam propagation in the atmosphere,'' \emph{Diala Jour}, vol.~36, pp. 2--9,
  2009.

\bibitem{JMLiuphotonic}
J.~M. Liu, \emph{Semiconductor lasers and light emitting diodes}, 1st~ed.\hskip
  1em plus 0.5em minus 0.4em\relax Photonic Devices, 2005.

\bibitem{edouard2013mathematical}
M.~Edouard and D.~Njomo, ``Mathematical modeling and digital simulation of {PV}
  solar panel using {M}atlab software,'' \emph{International Journal of
  Emerging Technology and Advanced Engineering}, vol.~3, no.~9, Sept. 2013.

\bibitem{aziz2014simulation}
M.~S. Aziz, S.~Ahmad, I.~Hassan, and U.~Saleem, ``Simulation and experimental
  investigation of the characteristics of a {PV}-harvester under different
  conditions,'' in \emph{2014 International Conference on Energy Systems and
  Policies (ICESP)}, Nov. 2014, pp. 1--8.

\bibitem{ding2009ofdm}
G.~Ding, Q.~Wu, N.~Min, and G.~Zhou, ``Adaptive resource allocation in
  multi-user ofdm-based cognitive radio systems,'' in \emph{2009 International
  Conference on Wireless Communications \& Signal Processing}, 2009.

\bibitem{cleveland2008battery}
T.~Cleveland, ``Battery charger adapts to multiple chemistries,'' \emph{Power
  Electronics}, vol.~34, no.~7, pp. 26--31, Jul. 2008.

\bibitem{dearborn2005charging}
S.~Dearborn, ``Charging {L}i-ion batteries for maximum run times,'' \emph{Power
  Electronics Technology}, vol.~31, no.~4, pp. 40--49, Apr. 2005.

\bibitem{chargingprofile}
Y.~S. Hwang, S.~C. Wang, F.~C. Yang, and J.~J. Chen, ``New compact {CMOS}
  {L}i-ion battery charger using charge-pump technique for portable
  applications,'' \emph{IEEE Transactions on circuits and systems}, vol.~54,
  no.~4, pp. 705--712, Apr. 2007.

\bibitem{dancy2000high}
A.~Dancy, R.~Amirtharajah, and A.~Chandrakasan, ``High-efficiency
  multiple-output {DC}-{DC} conversion for low-voltage systems,'' \emph{IEEE
  Transactions on Very Large Scale Integration (VLSI) Systems}, vol.~8, no.~3,
  pp. 252--263, Jun. 2000.

\bibitem{walker2004cascaded}
G.~R. Walker and P.~C. Sernia, ``Cascaded {DC}-{DC} converter connection of
  photovoltaic modules,'' \emph{IEEE Transactions on power electronics},
  vol.~19, no.~4, pp. 1130--1139, Jul. 2004.

\bibitem{810nmtransmitter}
L.~D. S. C. .~M. Technologies, ``Laser diode source - 808 nm,'' [Onilne].
  Available:
  \url{https://www.laserdiodesource.com/shop/808nm-25Watt-Laser-Diode-Module-BWT-Beijing}.

\bibitem{1550nmtransmitter}
L.~D.~S. Seminex, ``Laser diode source - 1550 nm,'' [Onilne]. Available:
  \url{https://www.laserdiodesource.com/laser-diode-product-page/1470nm-1532nm-1550nm-50W-multi-chip-fiber-coupled-module-Seminex}.

\bibitem{foghaze}
I.~I. Kim, B.~McArthur, and E.~Korevaar, ``Comparison of laser beam propagation
  at 785 nm and 1550 nm in fog and haze for optical wireless communications,''
  in \emph{Information Technologies 2000}.\hskip 1em plus 0.5em minus
  0.4em\relax International Society for Optics and Photonics, Feb. 2001, pp.
  26--37.

\bibitem{uniquempp}
N.~Femia, G.Petrone, G.~Spagnuolo, and M.~Vitelli, ``Optimization of perturb
  and observe maximum power point tracking method,'' \emph{IEEE transactions on
  power electronics}, vol.~20, no.~4, pp. 963--973, Jul. 2005.

\bibitem{dcenergyloss}
F.~Peng, F.~Zhang, and Z.~Qian, ``A magnetic-less {DC}-{DC} converter for dual
  voltage automotive systems,'' \emph{IEEE transactions on industry
  applications}, vol.~39, no.~2, pp. 511--518, Mar. 2003.

\end{thebibliography}
}

\end{document}